# Dimensional crossover and local strain induced deflection of the spin spiral state in multiferroic $NiI_2$

Tianxing Jiang,[1, §] Lianchuang Li,[1, §] Haiyan Zhu,[1, §] Hongyu Wang,[1] Junchao Tian,[1] Wenzhao Wang,[1] Weiyi Pan,[5] Haitao Wang,[1] Changlin Zheng,[1] Hongjun Xiang,[1,2] Changsong Xu,[1,2,4*] Donglai Feng,[3*] and Tong Zhang [1,4,6*]

[1]Department of Physics, State Key Laboratory of Surface Physics and Advanced Material Laboratory, Fudan University, Shanghai 200438, China

[2] Key Laboratory of Computational Physical Sciences (Ministry of Education), Institute of Computational Physical Sciences, Fudan University, Shanghai 200433, China

[3] New Cornerstone Laboratory, Hefei National Laboratory, Hefei 230088, China

[4] Hefei National Laboratory, Hefei 230088, China

[5] State Key Laboratory of Low Dimensional Quantum Physics and Department of Physics, Tsinghua University, Beijing 100084, China

[6] Shanghai Research Center for Quantum Sciences, Shanghai 201315, China

[§] These authors contributed equally to this work

* Corresponding authors: E-mail: csxu@fudan.edu.cn, dlfeng@hfnl.cn, tzhang18@fudan.edu.cn

**Low-dimensional multiferroics hold great promise for integrated magnetoelectric devices. Spin spiral state has recently been shown to induce ferroelectricity in single-layer van der Waals (vdW) material $NiI_2$. However, how this state evolves and can be tuned towards the two-dimensional limit remain unclear. Here, we combine spin-polarized scanning tunneling microscopy, layer-by-layer film growth, and multi-scale theoretical modeling to investigate the spin spirals in $NiI_2$ thin films. As the film thickness increases from 1 to 7 monolayers, we observed a continuous increase of spin-spiral wavelength and a rotation of wavevector from near [110] to [1$\bar{1}$0] direction, which evidences a dimensional crossover primarily driven by enhanced interlayer exchange energy. Moreover, we find that the film wrinkles can cause deflection of the spin spiral wavevector, which is caused by local curvature induced modification of exchange interactions. Our findings establish thickness and local strain as two tuning methods for engineering non-collinear helical magnetism and accompanied electric polarization in vdW multiferroics.**

## 1. Introduction

The emergence of layered van der Waals (vdW) magnets has opened a new frontier for exploring exotic magnetic states and designing integrated devices towards the two-dimensional (2D) limit [1-4]. Among them, the noncollinear magnets which break inversion symmetry, such as spin spirals, are of particular interest as they can induce ferroelectric polarization, leading to type-II multiferroicity [5-9]. The intrinsically strong magnetoelectric coupling in these spin-driven multiferroicity will promote potential applications in spintronics and nanoscale multi-functional devices. Consequently, elucidating the formation mechanisms and achieving precise control over these spin spirals in 2D systems have become critically important.

In general, spin spirals and related noncollinear textures can arise either from Dzyaloshinskii-Moriya interaction in non-centrosymmetric systems, or from competing exchange interactions in magnetically frustrated lattices [10-17]. In the latter case, spiral states are primarily stabilized by Heisenberg exchange couplings in combined with anisotropic interactions, such as the Kitaev and biquadratic terms [12-16]. The delicate balance of these interactions makes the noncollinear spin textures highly tunable via structural perturbations. In layered vdW magnets, these interactions are expected to evolve as the system approaches the 2D limit or is subjected to local deformations and strains [18-25]. However, systematic experimental studies on how thickness and local strain influence these microscopic mechanisms remain scarce.

$NiI_2$ is a prototypical vdW magnet hosting a spin spiral state [26]. Recently, robust type-II multiferroicity was observed in few-layer and monolayer $NiI_2$ [27-34]. Interestingly, the spin-spiral wavevector ($\boldsymbol{q}$) of monolayer $NiI_2$ differs significantly from that of bulk $NiI_2$ [34–36]. Although a realistic spin Hamiltonian incorporating frustrated Heisenberg and Kitaev terms has successfully explained the spin spiral in bulk $NiI_2$ [37], it cannot immediately explain the different wavevector of monolayer $NiI_2$. In addition, $NiI_2$ has been proposed to realize *p*-wave magnetism, in which intriguing odd-parity spin splitting energy bands are directly induced by the chiral spin spiral [38]. Therefore, investigating how the spin spiral in $NiI_2$ evolves with dimensionality, and how it can be effectively tuned, is important not only for understanding its microscopic origin, but also for precise design of its functional properties and device applications.

In this work, we systematically investigate the spin spiral state in few-layer $NiI_2$ films [1–7 monolayer (ML)] using spin-polarized scanning tunneling microscopy (SP-STM), combining with theoretical modeling and simulations. We found two different mechanisms which effectively tune the spin spiral state. First, we observe a continuous rotation of wavevector $\boldsymbol{q}$ and increase of wavelength as the film thickness increases, demonstrating a layer-resolved dimensional crossover. This behavior is attributed to enhanced interlayer exchange coupling that progressively rebalances the competition among the frustrated intralayer $J_1$–$J_3$–$K$ interactions. Second, we found the spin spiral state undergoes sharp reorientation at nanoscale wrinkles of the film. Our calculations show that it arises from local curvature induced modification of intralayer $J_3/J_1$ interactions. Overall, our findings establish film thickness and local topography as effective tuning knobs for vdW multiferroics, offering a practical route toward mechanically programmable and reconfigurable spin textures for future spintronic applications.

## 2. Materials and methods

### 2.1. Sample Preparation:

Few-layer $NiI_2$ film were grown by molecular beam epitaxy (MBE) on highly oriented pyrolytic graphite (HOPG) and graphitized SiC(0001) substrates. The HOPG was cleaved and outgassed at 400°C under high vacuum. Graphitized SiC(0001) was prepared by annealing SiC at 900°C under Si flux, followed by graphitization at 1350 °C. A single source of $NiI_2$ (99.5% purity powders) was used for film growth. The source was evaporated from Knudsen cell at 410 °C, while the substrate was kept at 140°C. The iodine background pressure during growth is ~ $4 \times 10^{-6}$ mbar, and the growth rate is about 20 – 30 min/ML. After growth, the samples were transferred *in situ* into the STM chamber.

### 2.2. SP-STM measurements

Spin-polarized STM experiment was conducted on the as-grown $NiI_2$ surface in a cryogenic STM system with a single-axis magnet (11 T along $\boldsymbol{Z}$) at $T$ = 4.5 K. Spin-resolved $\mathrm{d}I/\mathrm{d}V$ spectra and maps were acquired using Fe-coated tips, prepared by depositing 8−10 nm Fe film on an electrochemically etched W tip. Before Fe coating, the W tip was flashed up to ~2000 K to remove surface oxide. During the measurement, the Fe-coated tips were magnetized by applying $B_Z$ = 1T. Pure spin maps with spin sensitivity along $\boldsymbol{Z}$ ($S_Z$ map) were obtained from relative difference of $\mathrm{d}I/\mathrm{d}V$ maps acquired with tip magnetization along $+Z$ and $-Z$: $S_Z = (\frac{dI}{dV}|_{+Z} - \frac{dI}{dV}|_{-Z}) / (\frac{dI}{dV}|_{+Z} + \frac{dI}{dV}|_{-Z})$. Meanwhile, the charge map without spin signal were obtained by the sum of $\mathrm{d}I/\mathrm{d}V$ maps under $+Z$ and $-Z$ field. The $\mathrm{d}I/\mathrm{d}V$ spectra were collected by standard lock-in method at 741 Hz and the bias voltage ($V_\mathrm{b}$) is applied to the sample. To obtain spin contrast, $\mathrm{d}I/\mathrm{d}V$ maps were measured at energies within the conduction band of $NiI_2$, typically ranging from 1.0 to 1.8 eV, depending on the band position of different layers (see details in Part I-1 of Supplementary Material).

### 2.3. Scanning transmission electron microscopy (STEM) measurements

The cross-sectional atomic structure of $NiI_2$ was characterized using High-angle annular dark-field (HAADF) scanning transmission electron microscopy (STEM) imaging with an aberration-corrected STEM (Themis Z, Thermo Fisher Scientific) operated at 300 kV. The STEM samples were prepared with focused ion beam milling. HAADF-STEM images were collected with a probe semi-convergence angle of 21.4 mrad and detector collection angles ranging from 49 mrad to 200 mrad. To minimize sample drift and enhance the signal-to-noise ratio, a series of fast-scanning STEM images was acquired with a 500 ns pixel dwell time. The final HAADF-STEM image was then obtained by aligning and averaging 20 fast-scanning frames using an auto-correlation algorithm.

### 2.4. The density functional theory (DFT) calculation

In this work, DFT calculations were performed to generate and label a spin-lattice dataset for training a machine-learning potential of monolayer $NiI_2$, and to extract interlayer magnetic interaction parameters in few-layer $NiI_2$ for the subsequent analytical modeling. All DFT calculations were carried out using the Vienna Ab Initio Simulation Package (VASP) with the Projector Augmented Wave (PAW) method [39, 40] and the Perdew–Burke–Ernzerhof (PBE) functional [41]. A Hubbard correction ($U = 4.0$ eV) is applied to account for the 3d orbitals of Ni within the simplified rotationally invariant DFT+U approach proposed by Dudarev et al. [42]. Exchange parameters with different U values are shown in Part II-1 of Supplementary Material. A consistent plane-wave energy cutoff of 400 eV is used throughout the study.

DFT for dataset generation and labeling: For the primitive cell of monolayer $NiI_2$, a 17 × 17 × 1 Monkhorst–Pack k-point grid is used, with total energy convergence set to $10^{-7}$ eV per supercell. For larger 3 × 4 × 1 supercells in the dataset, atomic configurations are generated through several NVT molecular dynamics simulations. To avoid interlayer interactions, a vacuum buffer of 15 Å is implemented in the non-periodic axis. Static spin-polarized calculations, including spin–orbit coupling (SOC), are carried out for each selected atomic structure with ten spin configurations. Constraint DFT is applied to study systems with specified magnetic configurations using a 3 × 3 × 1 Monkhorst–Pack k-point grid, and convergence is set to $10^{-6}$ eV per supercell.

DFT calculation of interlayer interaction parameters: For the primitive cell of few-layer $NiI_2$ with AA stacking, a 4×4×3 supercell with a 3×3×2 Γ-centered k-point mesh is used. A reduced 2×2×2 Γ-centered k-point mesh was also tested, and the results showed no impact on the conclusions.

### 2.5. Spin–lattice dynamics simulations

Spin–lattice dynamics simulations are performed using a preconfigured LAMMPS module within the SpinGNN++ framework for spin optimization. Temperature is controlled by the Nosé–Hoover thermostat with a damping parameter of 100 time steps, whereas pressure is regulated by the Parrinello–Rahman barostat with a damping parameter of 1000 time steps. To model temperature-driven magnetic transitions, constant-volume (NVT) simulations are carried out to evolve the coupled atomic and spin degrees of freedom, incorporating a damping coefficient of 0.1 in the Landau–Lifshitz–Gilbert (LLG) equation. The magnetic moments of Ni are fixed at 2 $\mu_B$ throughout the simulations. For $NiI_2$, the initial random spin configuration is pre-relaxed using a conjugate-gradient (CG) algorithm for 1000 steps. Subsequently, constant-pressure spin–lattice dynamics (NPT–SIB) annealing is performed: the system is heated to 100 K over 15 ps, cooled to 0.01 K over a further 15 ps, and equilibrated for 5 ps. Finally, the spin configuration is relaxed using CG for 10,000 steps to obtain the final magnetic state.

## 3. Results

### 3.1 Surface structure characterization of $NiI_2$ film.

As shown in Fig. 1a and b, $NiI_2$ is a layered vdW material with a rhombohedral ($R\overline{3}m$) structure, where Ni atoms form a triangular lattice with an in-plane constant of 0.39 nm. The Ni $3d$ orbitals split into half-occupied $e_g$ and fully occupied $t_{2g}$ states, producing a local moment of ~$2\mu_B$ per Ni atom [37, 43-45]. We fabricated $NiI_2$ thin films via MBE on HOPG or graphene/SiC(0001) substrates, ensuring atomically flat surfaces (**Materials and methods**). The as-grown $NiI_2$ films show large scale terraces with varying thicknesses (1–7 ML, Fig. 1c). The atomic step height is ~0.63 nm, consistent with the thickness of a single $NiI_2$ monolayer (Fig. 1d). The layer numbers can be determined by its apparent height and thickness-dependent tunneling spectra, as detailed in Supplementary Material Part I-1. The d$I$/d$V$ spectra reveal a robust insulating gap of 2.1–2.2 eV in all the terraces of $NiI_2$ film (see Part I-1 of Supplementary Material).

The $NiI_2$ terraces are predominantly triangle shaped, with straight edges oriented along [100] and equivalent directions. The preference for a triangular (rather than hexagonal) terraces arises because the two types of edges shown in Fig. 1a and b have different atomic structures with respect to the substrate; therefore, the one with lower energy is preferred during MBE growth [35, 46-48]. In most cases, a series of triangular terraces with increased thicknesses have the same orientation, such as the terraces marked by 1–5 ML in Fig. 1c. This indicates the stacking order across these layers is identical. However, we can also occasionally observe terraces with an opposite orientation to the underneath layer (labelled by "mirror domain" in Fig. 1c). Atomically resolved STM image shows that a twin boundary is formed between the opposite domains (see Part II-2 of Supplementary Material). These observations indicate there are two different stacking configurations between two adjacent $NiI_2$ layers in the MBE-grown film.

To further verify the interlayer stacking order, we performed STEM on the cross section of $NiI_2$ film (**Materials and methods**). The atomic-resolved HAADF images clearly resolved two different stacking orders, as shown in Fig. 1e and f. The atomic structure of these stacking orders, referred as AA and AB', are sketched in Fig. 1b and superimposed on Fig. 1e and f. For AA stacking, the lattices of adjacent $NiI_2$ layers have the same orientation and are aligned vertically (no in-plane offset); while for AB' stacking the lattices are mirrored and Ni atoms in the upper layer align with the top I atoms of the lower layer. We note that both AA and AB' stacking are different from the AB stacking in bulk $NiI_2$, as shown in Fig. 1b (compared to AA stacking, the AB stacking has an interlayer shift along $[1\overline{1}0]$ direction). Via examining large scale STM images, we find that AA stacking is the dominant stacking order in our sample, as that marked in Fig. 1c [see also Fig. S3h (online)]. In the following we mainly focus on the magnetic structure of AA stacking regions, while the stacking order dependent magnetic structure is also an interesting topic and deserves further study [45, 49-53].

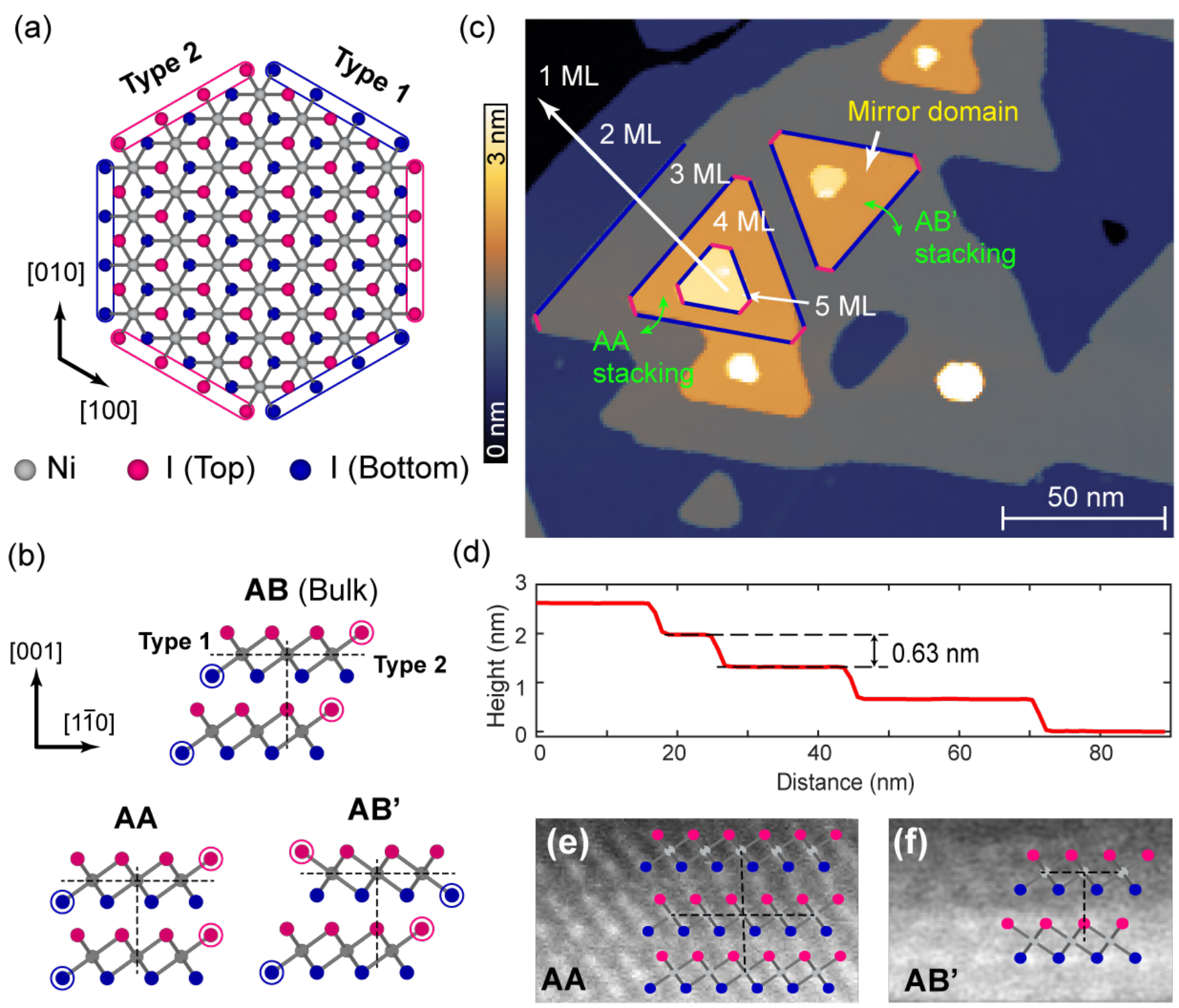

**Fig. 1. Lattice structure and STM/STEM characterization of few-layer $NiI_2$.** (**a**) The lattice structure of $NiI_2$ (top view). Two types of edges are marked with blue and red frame. (**b**) The lattice structure of $NiI_2$ with AB, AA and AB' stackings (side view). (**c**) Topographic image of the $NiI_2$ thin film sample ($V_b$ = 1.8 V, $I$ = 8 pA). Two types of edges are marked by blue/red lines. The terraces labelled by 1–5 ML have the same (AA) stacking order; while the terrace labeled by "mirror domain" has an AB' stacking with respect to underneath layer. (**d**) Line profile along the green arrow in (c), the single step height is 0.63 nm. (**e, f**) HAADF STEM images of the few-layer $NiI_2$ sample, showing the AA and AB' stacking, respectively.

### 3.2 Film thickness dependence of the spin spiral wavevector.

Then we performed spin-polarized measurements to investigate the spin structure. Fe-coated tips magnetized by out-of-plane magnetic fields are used (**Materials and methods**). On monolayer $NiI_2$, we observe stripe like patterns with a period of $\lambda$=1.74 nm (Fig. 2a). Its wavevector has an angle $\alpha \approx 7°$ relative to the [110] direction, yielding an incommensurate $\boldsymbol{q}$ = (0.13, 0.09, 0) in unit of reciprocal vectors (Fig. 2b). We note similar spin modulations have been reported in several prior STM studies of monolayer $NiI_2$ [30, 34-36], which are confirmed to be a spin spiral state with canted rotation plane [34]. Here for all the film thicknesses, stripe-like spin modulations are commonly observed. For example, Fig. 2c shows a pure spin contrast map with $S_Z$ resolution for 3ML terrace. The corresponding charge map, in which the spin signal is eliminated (**Materials and methods**), displays a 2$\boldsymbol{q}$ charge modulation similar to the 1ML case [34]. This indicates the common existence of spin-driven multiferroicity in $NiI_2$ few-layer films.

Fig. 2f shows an SP-STM image across a step edge. The spin spiral modulation exists in different terraces, while the phases of spin modulation between two adjacent terraces are nearly reversed. It implies an overall antiferromagnetic (AFM) like coupling between the neighboring Ni spins in two adjacent $NiI_2$ layers, which is consistent with the spin structure of bulk $NiI_2$ [26].

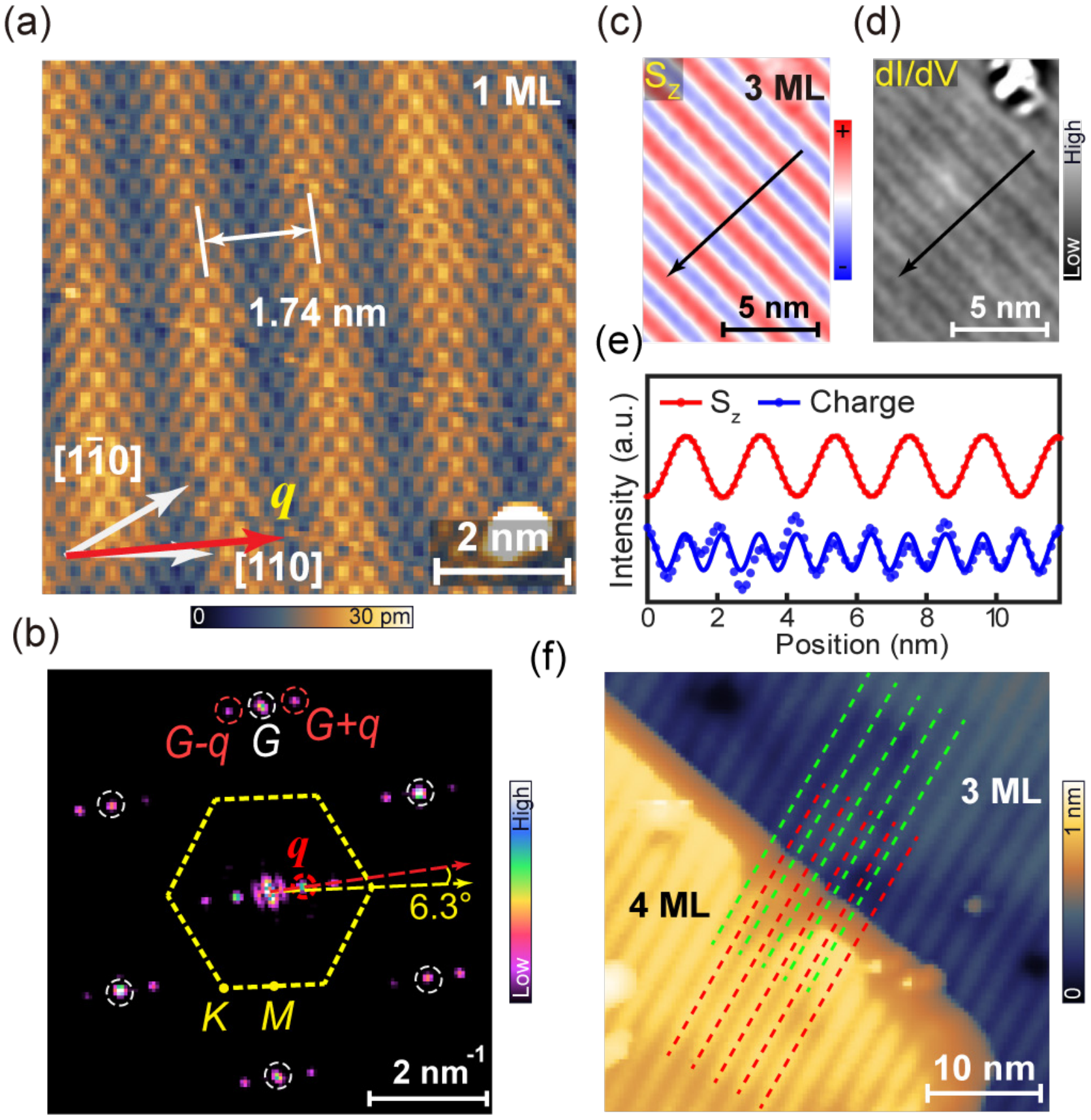

**Fig. 2. SP-STM measurement of spin spirals.** (**a, b**) SP-STM image of 1ML $NiI_2$ showing spin spirals stripes ($V_b$ = 1.0 V, $I$ = 80 pA) and its FFT image, respectively. (**c, d**) The $S_Z$ and charge maps taken on the same 3ML area, confirming magnetic contrast ($V_b$ = 1.35 V, $I$ = 50 pA). (**e**) The phase relationship between $\boldsymbol{q}$ and 2$\boldsymbol{q}$ along the arrow in Fig. 2c and d, further proved the spin spiral nature of stripes. (**f**) SP-STM image of a 3ML/4ML step ($V_b$ = 1.2V, $I$ = 30 pA). The spin spiral stripes display a half period offset across the step, evidencing an anti-parallel interlayer spin alignment.

As shown in Fig. 2a and reported in prior STM studies [34, 36], the spin spiral in 1ML $NiI_2$ significantly differs from that in bulk $NiI_2$, whose in-plane component of wavevector is along $[1\bar{1}0]$ direction [26, 29]. A key difference between 1ML and bulk $NiI_2$ is that interlayer interaction is completely absent in the monolayer, however this interaction would appear and becomes relevant as the film thickness increases. We therefore investigated the spin spiral in few-layer $NiI_2$ film with varying thicknesses. For each thickness, SP-STM images were acquired from 2–3 independent regions or terraces with lateral sizes of 10–50 nm. The spin-spiral wavevector, including the wavelength $\lambda$ and orientation angle $\alpha$, was extracted from the FFT of each image (see details in Part I-3 of Supplementary Material). Representative SP-STM images for different film thicknesses (1–7 ML) are shown in Fig. 3a–g. The spin-spiral wavelength ($\lambda$) and orientation of $\boldsymbol{q}$ are indicated in each image of Fig. 3a–g. As thickness increases, a general increase of $\lambda$ and rotation of $\boldsymbol{q}$ are observed. Quantitatively, $\lambda$ notably increases from 1.7 nm to 2.0 nm for 1–3 ML, and then converges to 2.1 nm for 7 ML, as summarized in Fig. 3h; while the angle between $\boldsymbol{q}$ and [110] direction (defined as $\alpha$) increases linearly from ≈7° for 1 ML to ≈26° for 7 ML (Fig. 3h). The gradual change of $\lambda$ and $\alpha$ with thickness suggests a crossover behavior when system evolves from single-layer to bulk limit (AA stacking). We note that the $\lambda$ of 7 ML film has not reached the value of bulk single crystal $NiI_2$ ($\lambda$ = 2.4 nm), while it already shows a saturated behavior. This could be due to the different interlayer stacking order in the MBE-grown film.

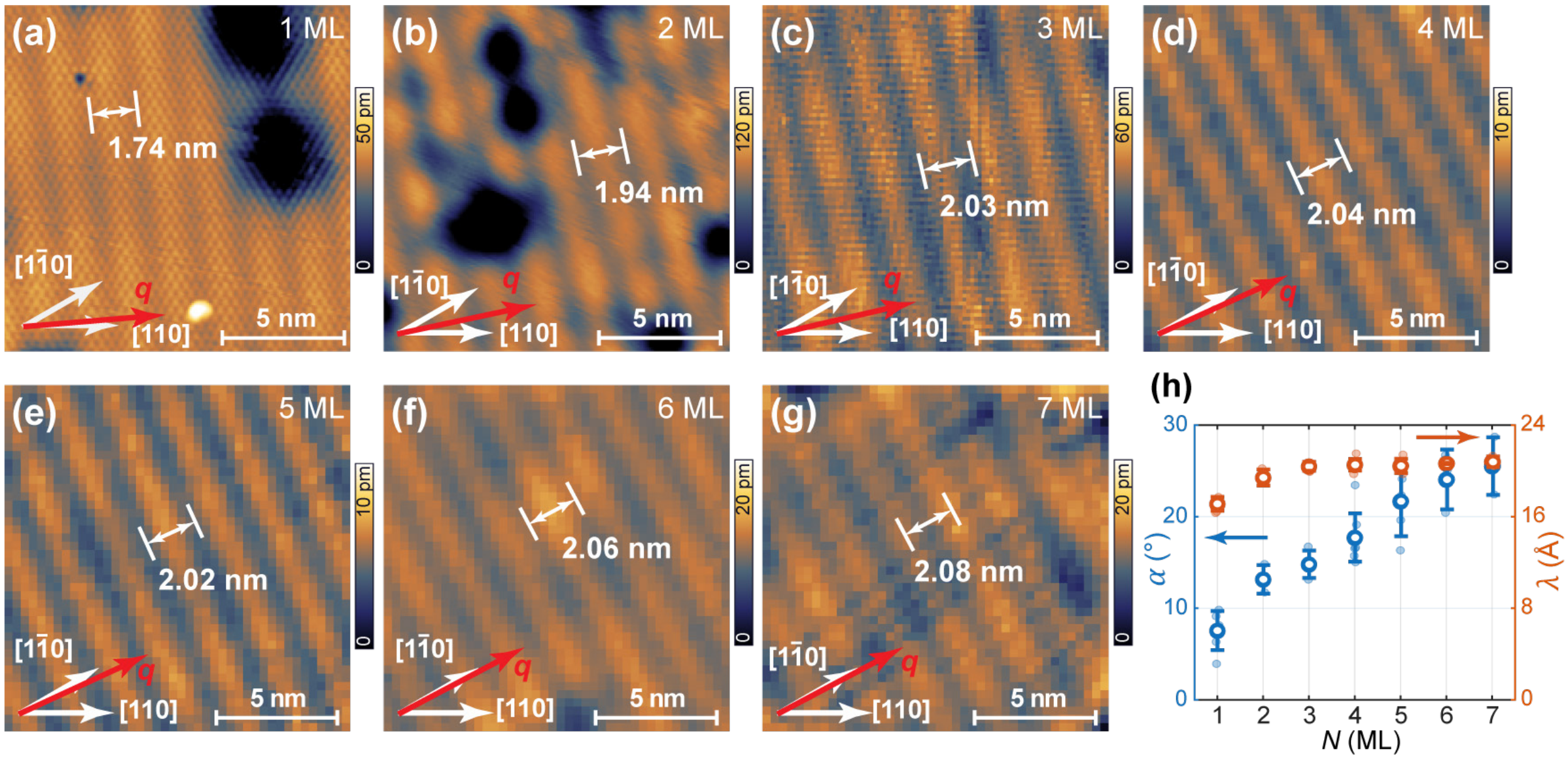

**Fig. 3. Thickness dependence of spin spirals.** (**a-g**) SP-STM images of the spin spirals at different thicknesses. The directions of $\boldsymbol{q}$ and $NiI_2$ lattice, and the modulation wavelength are marked in each image. (**h**) Relations of wavelength and direction vs. film thickness. Each light-colored data point corresponds to one independent measurement, the circles denote the corresponding mean values, and error bars indicate the standard deviations.

To elucidate the microscopic origin of varying spin spiral wavevector, we developed an analytical spin model for multi-layer AA-stacked $NiI_2$, based on a previous work [37, 54]. The spin Hamiltonian is defined as:

$$H = \sum_{\langle i,j\rangle_1} [J_1 \boldsymbol{S_i} \cdot \boldsymbol{S_j} + K S_i^\gamma S_j^\gamma + B(\boldsymbol{S_i} \cdot \boldsymbol{S_j})^2] + \sum_{\langle i,j\rangle_2} J_2 \boldsymbol{S_i} \cdot \boldsymbol{S_j} + \sum_{\langle i,j\rangle_3} J_3 \boldsymbol{S_i} \cdot \boldsymbol{S_j} + \sum_{\langle i,j\rangle_n^\perp} J_n^\perp \boldsymbol{S_i} \cdot \boldsymbol{S_j}$$

Here $J_1$ – $J_3$ are the 1st – 3rd nearest neighbor intralayer Heisenberg exchanges, $K$ is bond-directional Kitaev interaction, and $B$ is biquadratic interaction. For interlayer interactions, we consider Heisenberg exchanges $J_n^\perp$ between spins on adjacent layers in the AA stacking (see Fig. 4a for definition). For a single-$\boldsymbol{q}$ spin spiral state which is experimentally observed, the system's energy can be analytically derived from this Hamiltonian (see Part II of Supplementary Material for detail). Here the $B$ term could be absorbed into an effective $J_3$ term ($J_{3,\text{eff}} = J_3 + B/2$). The values of $J_1 - J_3$, $K$, $B$ for monolayer $NiI_2$ have been obtained from the previous work [37], and the values of $J_n^{\perp}$ in the AA stacking are calculated by DFT (see Part II-1 of Supplementary Material).

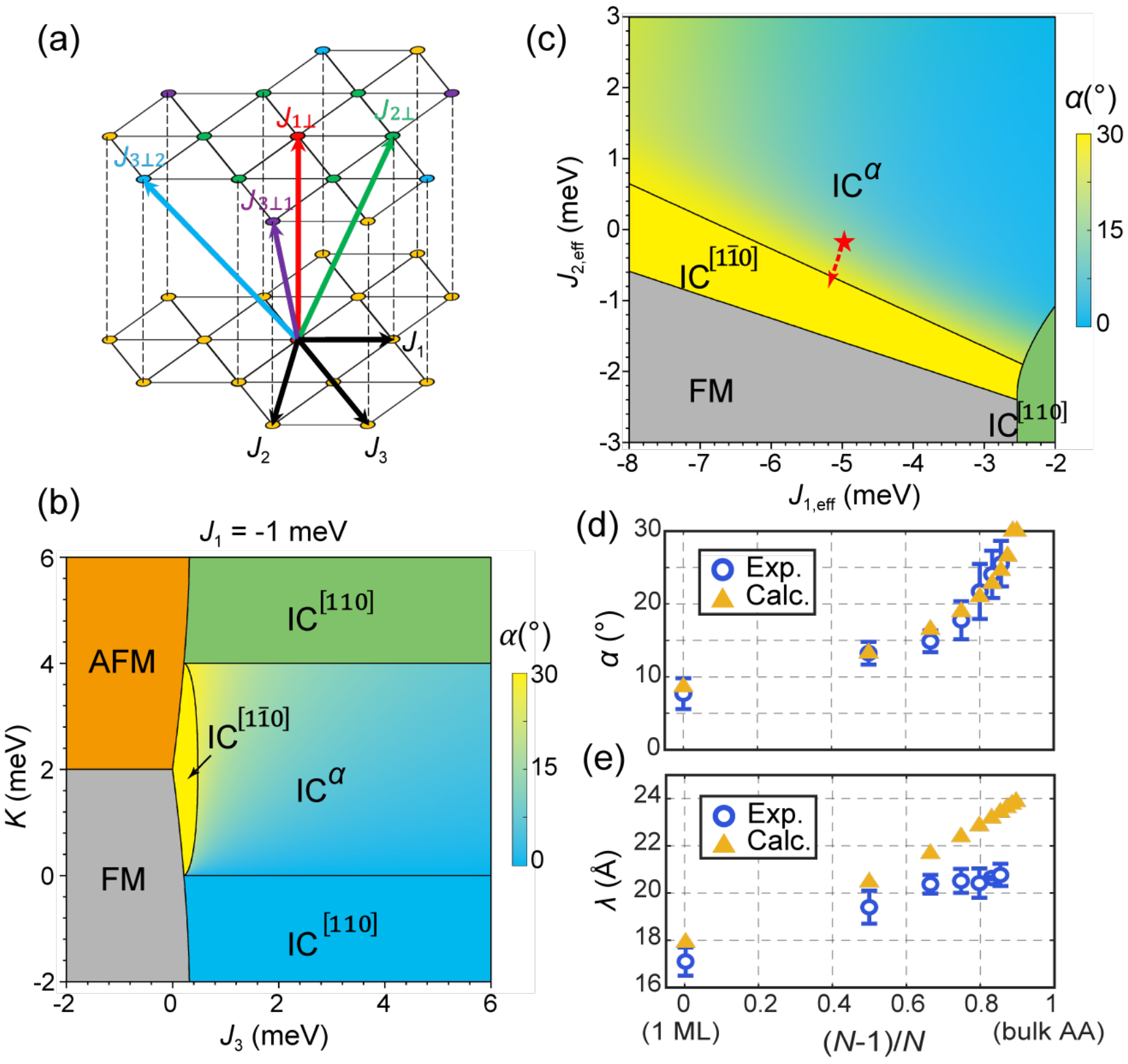

**Fig. 4. Phase diagram and Monte Carlo simulation.** (**a**) Illustration of the intralayer and interlayer Heisenberg exchange interactions (for AA stacking). (**b**) Phase diagram for 1 ML $NiI_2$ ($J_1$-$J_3$-$K$ model) in $J_3$-$K$ space. The false color of $IC^{\alpha}$ phase in the phase diagram represents the orientation angle ($\alpha$) of $\boldsymbol{q}$. The two $IC^{[110]}$ phases with blue/green color have different wavelength (see Part II-4 of Supplementary Material for more details). (**c**) Phase diagram of multi-layer $NiI_2$ (effective $J_1$-$J_2$-$J_3$-$K$ model) in $J_1$-$J_2$ space. Red star: the position of ($J_1$-$J_2$) for 1 ML $NiI_2$; Arrow: the trajectory towards bulk $NiI_2$ with AA stacking. (**d, e**) Comparison of experimental and calculation results of the $\boldsymbol{q}$ orientation and wavelength, respectively.

We first investigate the 1 ML system where the interlayer exchanges are absent. Since $J_1$ is relatively very small [37], the system can be described by a $J_1$-$J_3$-$K$ model. Through minimizing the energy for a single-$\boldsymbol{q}$ state with varying $\boldsymbol{q}$ in first Brillouin zone, the spin structure can be determined as a function of interaction parameters (see Part II-2 of Supplementary Material).

Fig. 4b shows a typical phase diagram in the ($J_1$, $K$) space (where $J_1$ is set as $-1$ meV for a referenced FM coupling). It is seen that for $J_3 > 0$ (AFM), the competition of these intralayer exchanges resulting an incommensurate (IC) spin spiral, whose wavevector $\boldsymbol{q}$ is sensitive to specific values of $J_3$ and $K$. Particularly, in the region marked by $IC^{\alpha}$ in Fig. 4b, the direction of $\boldsymbol{q}$ can vary continuously between [110] and $[1\bar{1}0]$ (indicated by the false color map); while in the regions marked by $IC^{[110]}$ and $IC^{[1\bar{1}0]}$, $\boldsymbol{q}$ is along [110] and $[1\bar{1}0]$, respectively. Therefore, a continuous variation of $\boldsymbol{q}$ is already reflected in the monolayer phase diagram.

We then incorporate the interlayer exchange couplings. As illustrated in Fig. 4a, the 1st – 3rd interlayer interactions (for AA stacking) are denoted as $J_1^{\perp}$, $J_2^{\perp}$, $J_{3,1}^{\perp}$, $J_{3,2}^{\perp}$. In Fig. 2f we have shown that the Ni spins between adjacent layers are AFM coupled (*i.e.*, $\boldsymbol{S}_{R_0} = -\boldsymbol{S}_{R_{0,\perp}}$). With this relation, the interlayer couplings in few-layer $NiI_2$ can be treated as modifications of intralayer couplings in the monolayer system. The resulting effective intralayer interactions are: $J_{1,\mathrm{eff}} = J_1 - \frac{2(\mathrm{N}-1)}{\mathrm{N}} J_2^{\perp}$, $J_{2,\mathrm{eff}} = J_2 - \frac{(\mathrm{N}-1)}{\mathrm{N}} J_{3,2}^{\perp}$, where $N$ is the film thickness. The coefficient $(N-1)/N$ appears because for a $N$-layer film the "averaged" intralayer interactions for each layer is proportional to $(N-1)/N$ (see Part II-5 of Supplementary Material for details). We thus include $J_{2,\mathrm{eff}}$ into the original model, by adopting DFT values of $J_3$ and $K$ [Table S1 (online)], a new phase diagram in the ($J_{1,\mathrm{eff}}$, $J_{2,\mathrm{eff}}$) space is shown in Fig. 4c. When $N$ increases, the $J_{1,\mathrm{eff}}$, $J_{2,\mathrm{eff}}$ values will vary along the path indicated in Fig. 4c, which demonstrates the dimensional crossover of the system.

In Fig. 4d and e we show the calculated $\boldsymbol{q}$ vectors as function of $N$ based on the above model. A good agreement is obtained between calculated orientation of $\boldsymbol{q}$ and the experimental data. Meanwhile the calculated wavelengths also capture the observed trend, while slight quantitative discrepancies remain. The difference could be due to that the measured $NiI_2$ terrace has finite size, which may alter effective magnetic interactions, while our model did not consider this effect.

We note that the above calculation is restricted to AA stacking order. Nevertheless, the actual AB-stacked structure also exhibits weak AFM interlayer interactions [37] (see Part II-6 of Supplementary Material for more discussion). Therefore, we expect similar dimensional crossover behavior to appear in AB stacking case, although the specific thickness-dependence may be different.

### 3.3 Film curvature induced deflections of spin spiral

We next examine another factor which makes spin spiral state changes locally. In the MBE grown $NiI_2$ films, some wrinkle-like structures can be observed, as shown in Fig. 5a. They are likely induced by the unevenness of substrate (e.g., the step edges) and the flexibility of the layered structure of $NiI_2$. Interestingly, the spin modulations, or wavevector of spin spiral, are found to be deflected when they cross the wrinkles, and recover normal orientation after leaving it (Fig. 5c). A similar behavior is also observed when the film crosses step edges of graphene/SiC substrate (where a locally bent region is generated), as shown in Fig. 5b and d. The local curvature ($\kappa$) of these wrinkles or bent regions can be extracted from the profile of topographic images after averaging along the wrinkle direction, which is in the range of 10–20 $\mu m^{-1}$ (Fig. 5e and f, see details in part II-3 of Supplementary Material). We have examined various samples and confirmed that such local curvature induced deflection of spin spiral is a common behavior (see Part I-4 of Supplementary Material for additional data). The deflection angle of spin spirals in curved regions is mainly distributed in the range of 40°–68°. We note that such deflected spin spiral creates a boundary-like region between spin spirals with different local $\boldsymbol{q}$-vectors. However, such boundary should be distinguished from the spin spiral domain wall reported before [34], where topological spin structures are formed due to the overlap of two spin spirals. In the present case, the deflection corresponds to a local reorientation of a single-$\boldsymbol{q}$ spiral driven by curvature-induced modification of magnetic interactions.

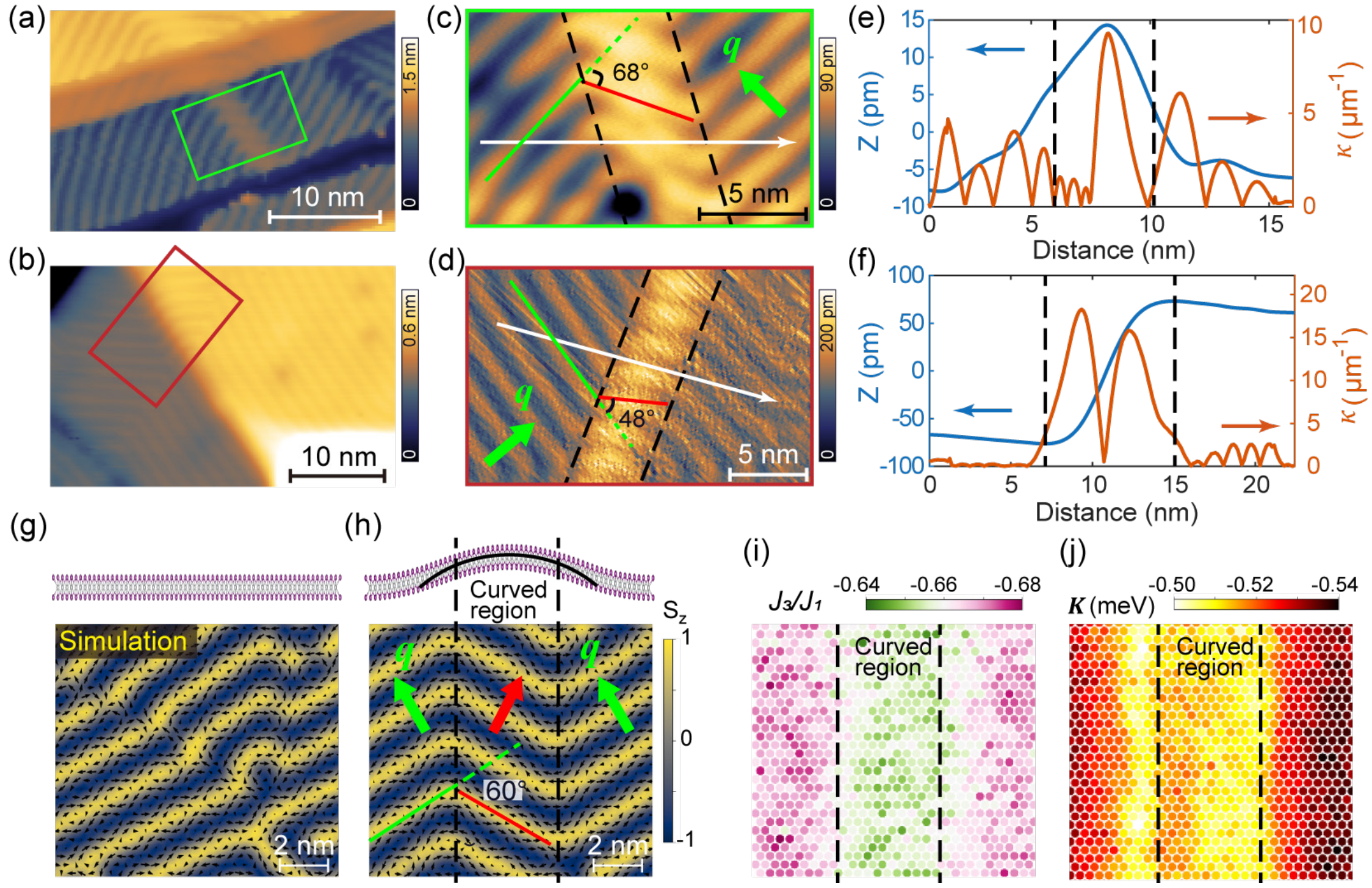

**Fig. 5. Curvature induced deflection of spin spirals.** (**a, b**) Large-scale STM image containing curved region. (**c, d**) Zoomed-in images of the marked regions in (a). The deflection angles of the spin spiral at curved areas are indicated. [panel (d) is shown as differential image] (**e, f**) The line profiles averaged along the vertical direction, taken along the white arrows in (c, d) and the calculated local curvature. (**g, h**) Simulated spin structures of a flat and a curved 1 ML $NiI_2$ film, respectively (the corresponding lattice models are shown on top of the images). Deflected spin spiral is observed in the curved region between two dashed lines in (h). The small kinks of the spin spiral in (g) arise from finite precision of the spin-lattice dynamics simulation where spin/lattice are relaxed simultaneously. (**i, j**) Spatial distributions of $J_3/J_1$ and Kitaev interaction ($K$) of the curved lattice same as (h), respectively.

To elucidate the microscopic origin of this local curvature-induced spin structure modification, we conducted large-scale spin-lattice dynamics simulations. Since standard DFT method is difficult to treat non-uniform strain in large supercells, we applied a machine-learning based SpinGNN++ framework to conduct the spin-lattice dynamics simulations[55]. In this approach, a neural network is employed to predict magnetic structures over large-scale curved lattices by learning the DFT-computed magnetic energy for different structural and spin configurations $(\boldsymbol{r}, \boldsymbol{s})$ (see details in Methods and Part III of Supplementary Material). We constructed a curved 1ML $NiI_2$ lattice by mapping it onto a cylindrical surface with a curvature of $\kappa \sim 10\ \mu m^{-1}$ (Fig. 5h, top), and then relaxed both atomic and spin degrees of freedom by spin-lattice dynamics. The bent geometry remains after relaxation, showing that the curved structure is a metastable configuration. Fig. 5h shows the simulated spin texture of the curved lattice, which exhibits a clear deflection of spin spirals (with an angle of ~ 60°) in curved region. This is qualitatively consistent with the experimental observation in Fig. 5c and d. For comparison, Fig. 5g shows the simulation of a non-curved lattice, where no such regular deflection is observed; the small kinks of spin spiral arise from finite precision of spin-lattice dynamics simulation.

Moreover, we can extract the spatial variation of Heisenberg exchange interactions and Kitaev interaction ($K$) through the SpinGNN++ framework (by predicting the magnetic exchange tensor)[55]. As shown in Fig. 5i and j, on the curved lattice, the $J_3/J_1$ ratio and $K$ show notable reduction of their absolute values in the curved region. Microscopically, the spin spiral in $NiI_2$ should be regarded as a spin–lattice coupled state. In flat regions, the lattice distortion associated with spin spiral is already incorporated into the equilibrium helical state. At wrinkles, additional nonuniform structural deformation is imposed locally, which further changes the Ni-I bond lengths and bond angles (the Ni-I bonds are compressed in the inner side and stretched on the outer side, as shown in Fig. S6 (online)) [19, 56]. Consistently, the SpinGNN++ analysis shows that the effective $J_1$, $J_3$, and $K$ are modified in the curved region. Since the spin-spiral wavevector is determined by a delicate balance among these competing interactions, such local structural perturbation can shift the preferred wavevector and lead to the observed deflection. The observed stripe deflection therefore manifests the local variation of the competing magnetic interactions, which is distinct from previously reported domain walls between different single-$\boldsymbol{q}$ spin spiral domains [34]. To examine this point, we calculated the topological charge density from the simulated spin configuration, as shown in Fig. S6f (online). No robust topological charge is found at the deflected region.

## 4. Discussion and Conclusion:

We have now identified two distinct mechanisms that can modify the spin-spiral wavevectors in few-layer $NiI_2$ films. The thickness dependence of $\boldsymbol{q}$ reveals how interlayer interactions reshape the spin spiral state. For 1ML $NiI_2$, the competing intralayer interactions of $J_1$, $J_3$, and $K$ determine the wavevector $\boldsymbol{q}$. Upon increasing the thickness $N$, AFM-like interlayer interaction $J_{\mathrm{n}}^{\perp}$ introduces effective FM in-plane interactions with a factor $(N\text{-}1)/N$, thereby shifting the system in $\mathrm{IC}^{\alpha}$ phase from near-$\mathrm{IC}^{[110]}$ to near-$\mathrm{IC}^{[1\bar{1}0]}$. This provides microscopic insight into how interlayer exchange modifies the frustrated magnetic ground state.

The curvature-induced deflection of spin spirals reveals a different mechanism, where curved surface perturbs the intralayer exchange interactions locally. The rotation symmetry-breaking wrinkles (and step edges) introduce variation of $J_1/J_3$ ratio and $K$ only in the curved region, thereby inducing deflection of the spin spiral stripes when they cross these structural features. This establishes few-layer $NiI_2$ as a model system for investigating how geometric perturbations can manipulate spin structure at the nanoscale. Moreover, the curvature-induced deflection of spin spiral may locally modify the associated electric polarization. Although the electric polarization is not directly measured by STM, the nanoscale modification of spin spiral provides a possible route for mechanically programming the multiferroic domains and directly writing ferroelectric polarization patterns.

Meanwhile, several important issues remain open for future investigation. Because the interlayer stacking order (AA) of $NiI_2$ MBE film is different from the bulk crystal, whether alternative stacking configurations exhibit different crossover pathways is worthy of further study. In addition, the interplay between the evolving spin spiral structure and electric polarization in few-layer systems remains largely unexplored, and understanding how the dimensional crossover influences magnetoelectric coupling will be of particular interest. It is also worth discussing the possible effect of interfacial charge transfer between the substrate and $NiI_2$ film. However, this effect is unlikely to dominate the observed dimensional crossover. As shown in Part I-5 of Supplementary Material, $NiI_2$ films grown on $NbSe_2$ substrate still exhibits spin-spiral wavevectors similar to those on HOPG, even though their $dI/dV$ spectra indicate significant difference in interfacial charge transfer. A similar phenomenon has also been reported for monolayer $NiI_2$ grown on Au [35]. These results suggest that substrate does not play a dominant role in the evolution of spin-spiral order in $NiI_2$ film.

In summary, by combining atomically resolved SP-STM and multi-scale theoretical modeling, we have systematically unveiled the effects of thickness and local strain on the spin spiral state in vdW multiferroic $NiI_2$. The continuous evolution of the spin spiral wavevector demonstrates layer-dependent dimensional crossover driven by interlayer interactions. Local reorientation of spin spirals at structural wrinkles uncovers a microscopic magneto-elastic coupling. Our findings establish film thickness and local strain as two accessible turning knobs for designing complex non-collinear magnetic textures. This framework not only deepens the fundamental understanding of frustrated magnetism in reduced dimensions but also paves a practical pathway for designing reconfigurable nanoscale magnetoelectric and spintronic devices.

## Conflict of interest:

The authors declare that they have no conflict of interest.

## Acknowledgments:

This work is supported by National Natural Science Foundation of China (12225403, 92365302, 12188101, 12174060, 12274082), Quantum Science and Technology-National Science and Technology Major Project (2021ZD0302803, 2024ZD0300102, 2024ZD0300103), the New Cornerstone Science Foundation, China (NCI202211), Shanghai Municipal Science and Technology Major Project (2019SHZDZX01), Shanghai Pilot Program for Basic Research (21TQ1400100), Xiaomi Young Talents Program.

## Author Contributions

Tianxing Jiang, Hongyu Wang, Junchao Tian, and Haitao Wang prepared the sample and performed STM measurement. Lianchuang Li, Haiyan Zhu, and Weiyi Pan performed the analytical modeling and calculation under the guidance of Changsong Xu and Hongjun Xiang. Wenzhao Wang and Changlin Zheng performed the STEM measurement. Donglai Feng, Changsong Xu, and Tong Zhang coordinated the project. Tianxing Jiang, Lianchuang Li, Haiyan Zhu, and Tong Zhang wrote the manuscript. All authors have discussed the results and the data interpretation.

## Appendix A. Supplementary material

Supplementary materials to this article can be found online at (link xxx).

# Supplementary Materials for

# Dimensional crossover and local strain induced deflection of the spin spiral state in multiferroic $NiI_2$

Tianxing Jiang,[1,§] Lianchuang Li,[1,§] Haiyan Zhu,[1,§] Hongyu Wang,[1] Junchao Tian,[1] Wenzhao Wang,[1] Weiyi Pan,[5] Haitao Wang,[1] Changlin Zheng,[1] Hongjun Xiang,[1,2] Changsong Xu,[1,2*] Donglai Feng,[3*] and Tong Zhang [1,4,6*]

[1]Department of Physics, State Key Laboratory of Surface Physics and Advanced Material Laboratory, Fudan University, Shanghai 200438, China

[2] Key Laboratory of Computational Physical Sciences (Ministry of Education), Institute of Computational Physical Sciences, Fudan University, Shanghai 200433, China

[3] New Cornerstone Laboratory, Hefei National Laboratory, Hefei 230088, China

[4] Hefei National Laboratory, Hefei 230088, China

[5] State Key Laboratory of Low Dimensional Quantum Physics and Department of Physics, Tsinghua University, Beijing 100084, China

[6] Shanghai Research Center for Quantum Sciences, Shanghai 201315, China

[§] These authors contributed equally to this work

* Corresponding authors. E-mail: csxu@fudan.edu.cn, dlfeng@hfnl.cn, tzhang18@fudan.edu.cn

## Part I: Additional STM data and data processing details

### I-1: Determination of the $NiI_2$ layer number and layer-dependent d*I*/d*V* spectra

In the presented STM measurements, the thickness of $NiI_2$ terraces was determined by combining topographic profile and the tunneling spectroscopy.

1) At low-coverage regions, as shown Fig. S1a for example, the exposed graphene or HOPG substrate can be easily identified in topographic images, as they display much less point defects than $NiI_2$ terraces. Once the substrate is identified, the thicknesses of adjacent $NiI_2$ terraces can be directly determined from their apparent height relative to substrate (Fig. S1b).

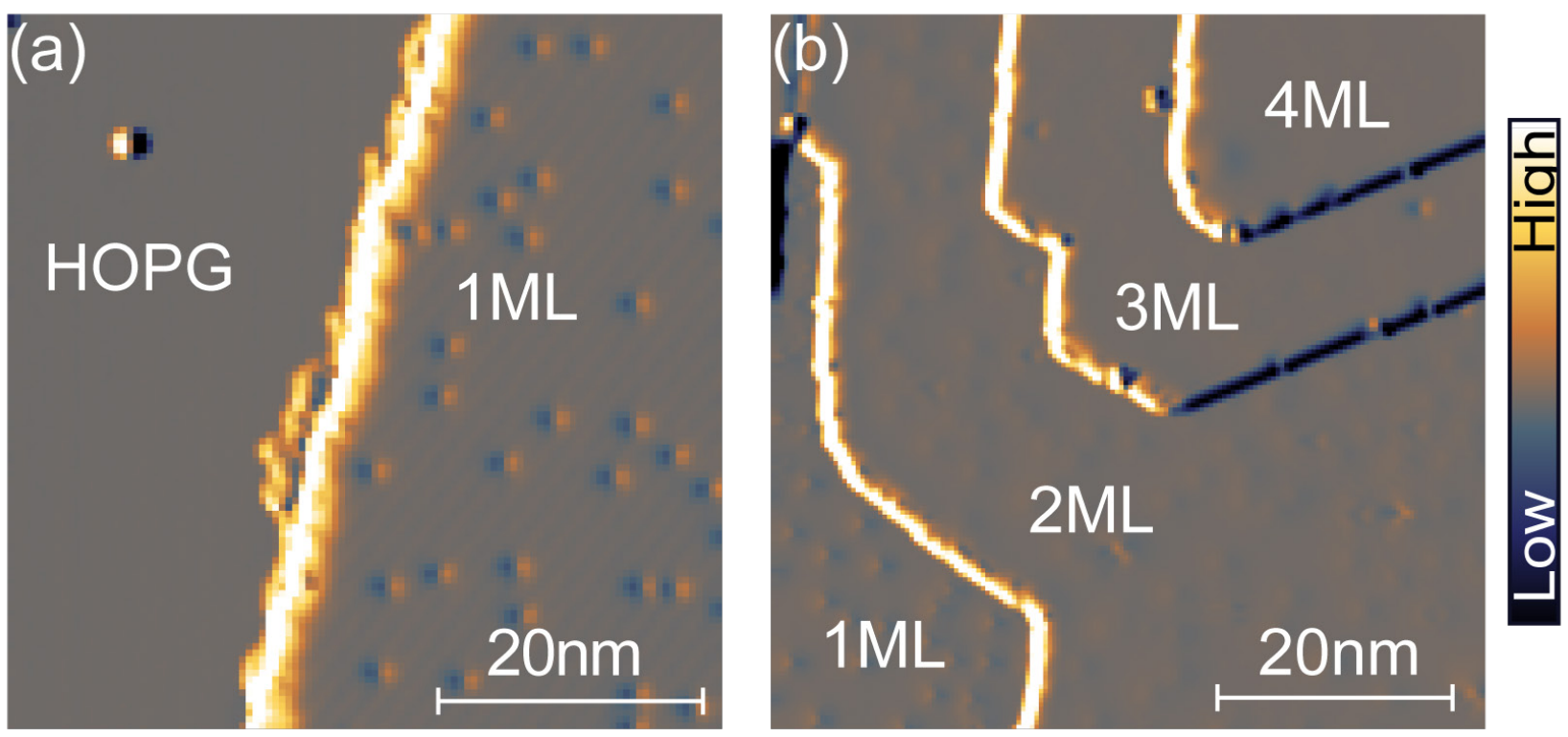

**Fig. S1. Determination of layer number.** (**a**) Topographic image (differential mode) of 1ML $NiI_2$ film and nearby HOPG substrate. (**b**) Topographic image (differential mode) of 1 - 4ML $NiI_2$ terraces.

2) The tunneling spectrum provides another independent criterion for thickness identification. As shown in Fig. S2a and b. The spectra display a large insulating gap of ~2 eV on all measured thicknesses. As thickness increases, the conduction band edge shifts to positive energy systematically, establishing a one-to-one correspondence between the spectral features and the film thickness. We note that although some spectra were acquired using spin-polarized tips in Fig. S2a, they mainly reflect the total local DOS of $NiI_2$, since the spin signal is considerably small (<10%). We note that for bulk $NiI_2$, the insulating gap is reported to be ~ 1.6–1.9 eV based on photoconductivity measurements [1]. The slightly large gap observed for $NiI_2$ thin film is likely due to reduced dimensionality and/or the different gap size probed by tunneling spectroscopy versus optical measurements.

We also performed DFT+U calculations on the DOS of isolated few-layer $NiI_2$. The calculated DOS shows a clear insulating gap for all thicknesses from 1 to 5 layers (Fig. S3). The gap only decreases slightly from monolayer to few-layer $NiI_2$. This weak thickness dependence is consistent with the experimental data in Fig. S2c that the large tunneling gap persists from monolayer to few-layer films.

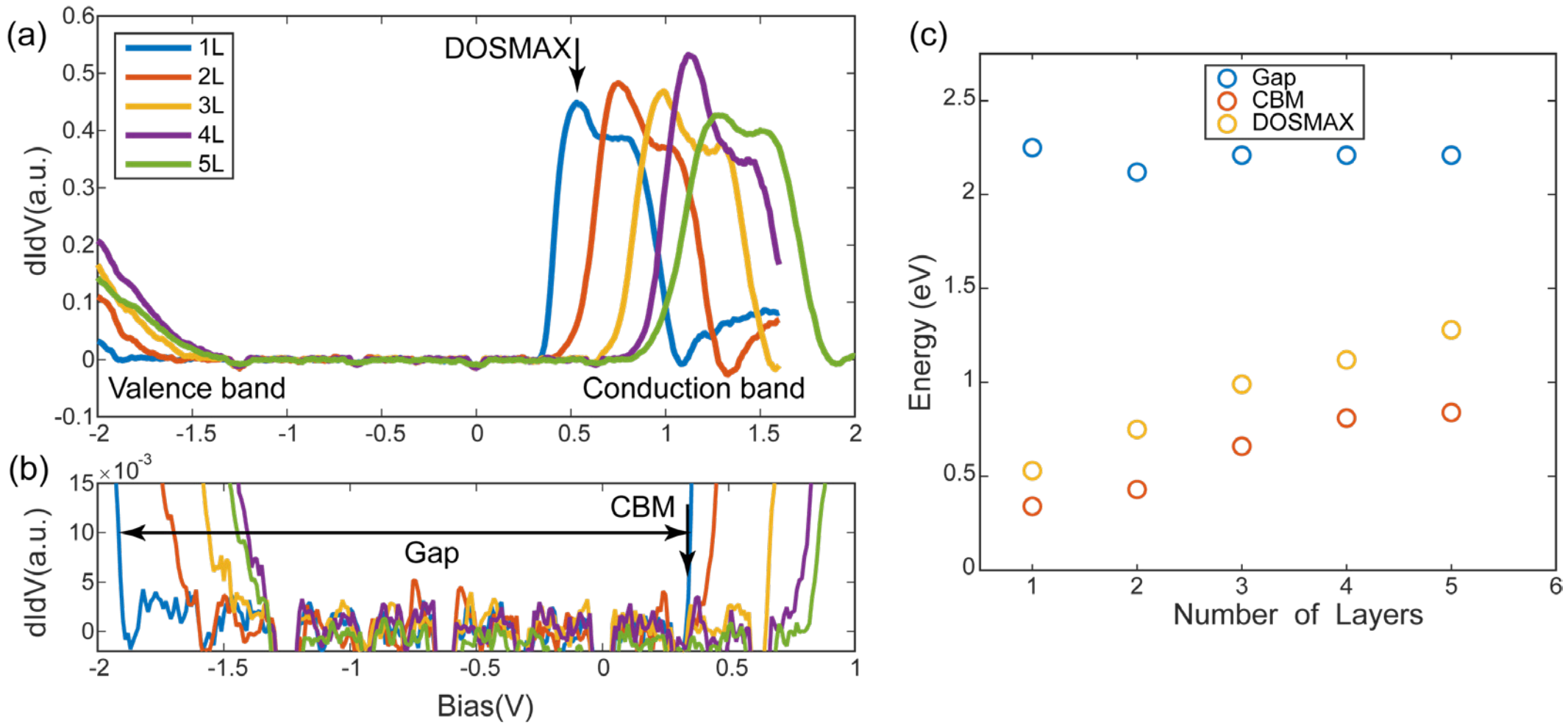

**Fig. S2.** (**a**) Tunneling spectra taken on $NiI_2$ film form 1 layer to 5 layers. (**b**) Zoom-in of spectra in (a), showing clear band moving behavior. (**c**) The bandgap size, conduction band minimum (CBM) and maximum of DOS (DOSMAX) as function of layer number.

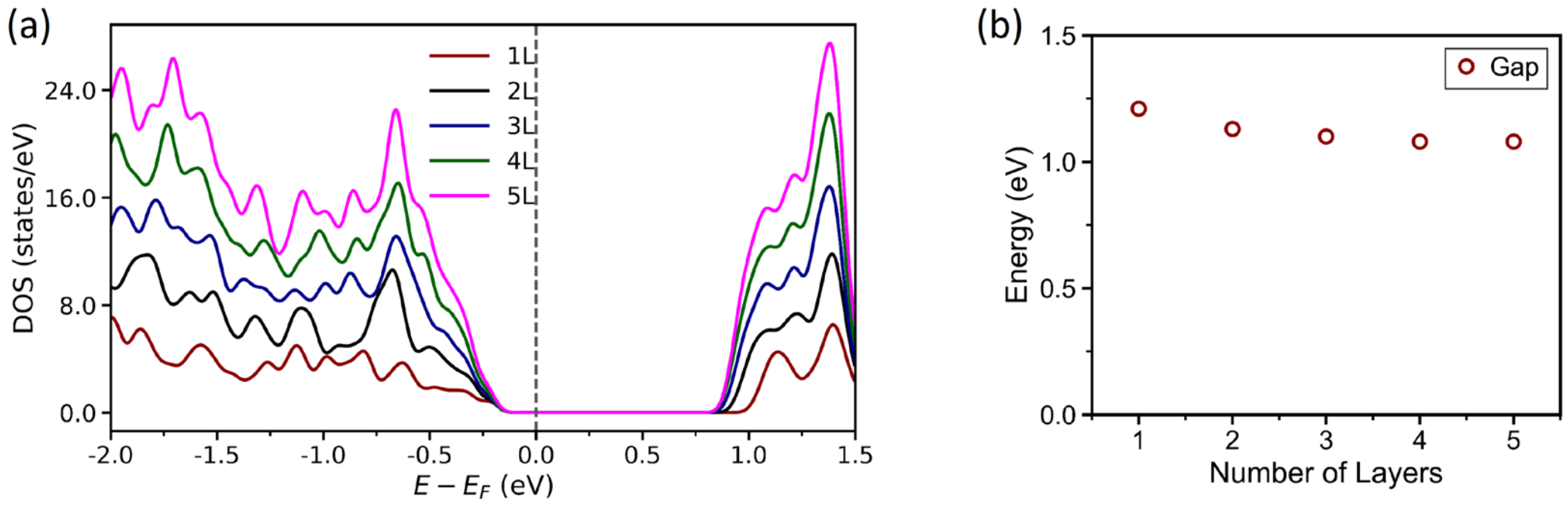

**Fig. S3.** (**a**) Total DOS of 1–5 layer $NiI_2$ calculated by DFT+U. The energy is aligned to the Fermi level of each layer. (**b**) Calculated band gap as a function of layer number, showing a weak thickness dependence from 1.21 eV for monolayer to 1.08 eV for 5 layers.

## I-2: Determination of the interlayer stacking order

Besides the STM image and STEM data shown in Fig. 1c and Fig. 1e and f of the main text, the two different stacking orders are further evidenced by the existence of twin boundaries. We observed that when islands with different orientation coalesce, a mirror-twin boundary (MTB) is formed (Fig. S4a), similar to that observed in $FeCl_2$ [2]. Atomically-resolved image (Fig. S4e) reveals a lattice offset across the boundary, consistent with the expected MTB structure shown in Fig. S4f. Meanwhile, the $NiI_2$ terraces with the same orientation can merge without any domain boundaries formed (Fig. S4b), and the lattices of adjacent layers have no offset (Fig. S4c and d), indicating an AA stacking. Therefore, the two sides of the MTB boundary should have different stacking orders, one is AA and another is

AB’, as illustrated in Fig. S4g.

From large-area image, the stacking mode of different $NiI_2$ triangular islands can be identified by the orientation. The statistics of stacking mode show that AA stacking mode is dominant in our samples comparing to AB’ stacking mode (Fig. S4h). Within the resolution of STM and HAADF-STEM data, we did not observe other types of stacking configurations beyond AA and AB’.

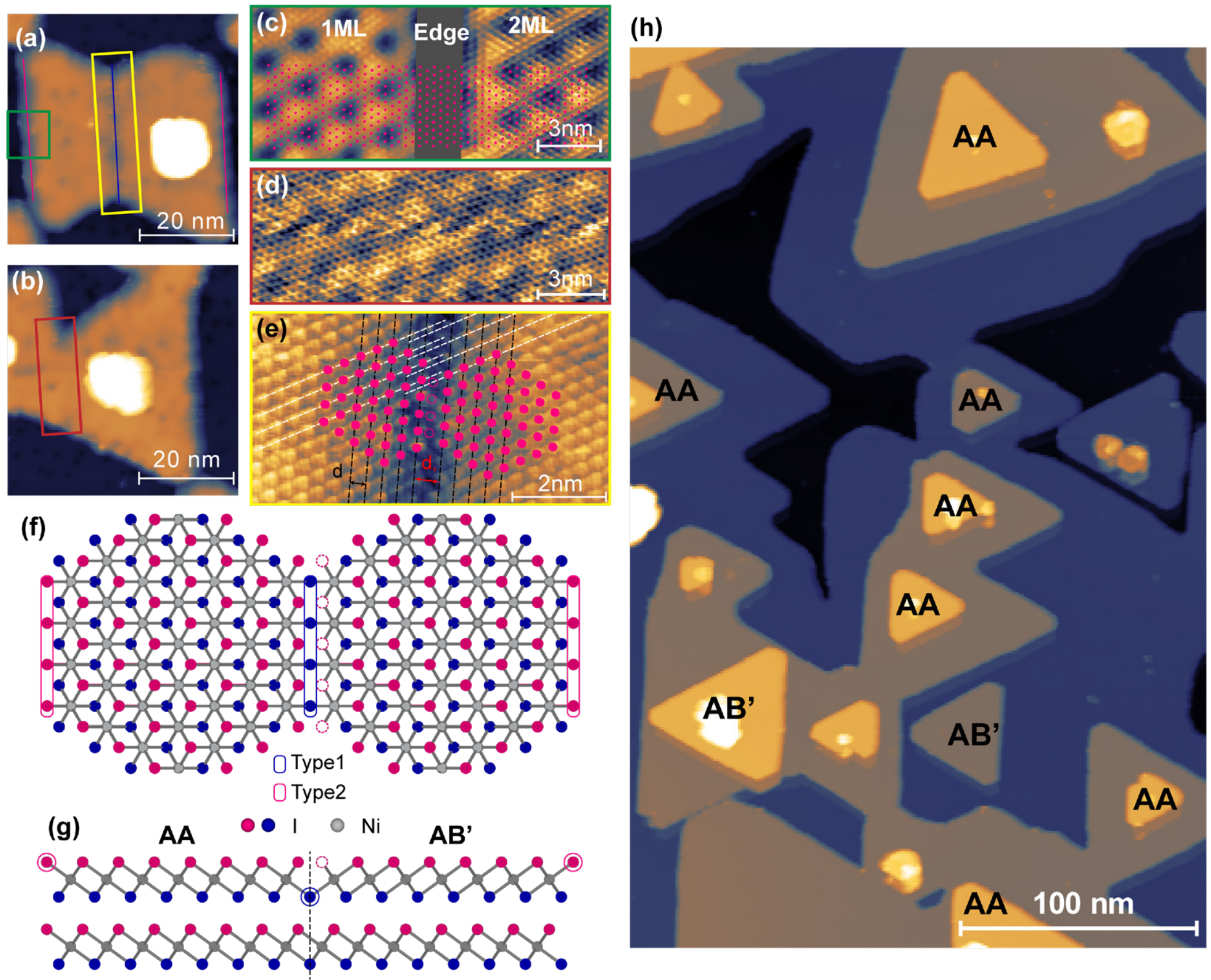

**Fig. S4.** (**a, b**) Topography image where triangular islands connect to each other. Blue and red lines marked the type of edges. (**c**) Atomic resolved image at edge in the green frame of (a). AA stacking can be resolved here. (**d**) Atomic resolved image at continuous boundary in the red frame of (b). (**e**) Atomically resolved image at mirror twin boundary in the yellow frame of (a). Lattice of top I atoms resolved here are staggered to each other between domains. (**f, g**) Vertical and front views of atomic structure of mirror twin boundary. (**h**) Large-scale STM image showing that the AA stacking is the dominating stacking.

### I-3: Data processing details about wavevector of spin spirals.

The spin-spiral wavevector was extracted from the two-dimensional fast Fourier transform (FFT) of the SP-STM images. For each measured area, the crystallographic lattice direction was determined by the atomic lattice or the edges of the film, and the spin-spiral wavevector $\boldsymbol{q}$ was then determined from the position of the magnetic modulation peak in the FFT. The spin-spiral wavelength was calculated as $\lambda = 2\pi/|\boldsymbol{q}|$, and the orientation angle α was defined as the angle between $\boldsymbol{q}$ and the [110] crystallographic direction. For each film thickness, independent SP-STM images acquired from different regions/terraces were analyzed separately. Each light-colored data point in Fig. 3h corresponds to one independent FFT analysis, while the solid symbol represents the mean value for that thickness. The error bars denote the standard deviation of the individual measurements. The same statistical treatment was used for the experimental data shown in Fig. 4d and e.

**I-4: Additional data and analysis on local curvature induced spin spiral deflection.**

We investigated various wrinkles in different regions of the sample, and confirmed the common existence of this curvature induced phenomenon (Fig. S5). Although the formation and detailed geometries may vary in different regions, the same qualitative behavior is repeatedly observed: the spin-spiral stripes change their direction upon crossing a wrinkle oriented at a finite angle relative to them (Fig. S5a–f). The deflection angle of spin spirals is in the range of 40–60°. while the stripes keep unchanged if they are parallel to the wrinkle (Fig. S5g and h).

To quantify the local curvature, the STM topographic profile was first averaged along the wrinkle direction within a finite-width region, and the resulting averaged height profile $z(x)$ across the wrinkle was used to calculate the local curvature by $\kappa(x) = |d^2 z/dx^2|/[1 + (dz/dx)^2]^{3/2}$. The curvature of different regions is generally on the order of $\kappa \sim 10\ \mu\mathrm{m}^{-1}$.

Fig. S5i shows a pure spin contrast $S_Z$ map of a curved region. It's seen that the spin modulation displays abrupt deflection rather than the superposition of two spin spiral states. This is markedly different from the topological spin textures observed at $NiI_2$ spin-spiral domain walls[3], indicating a different formation mechanism.

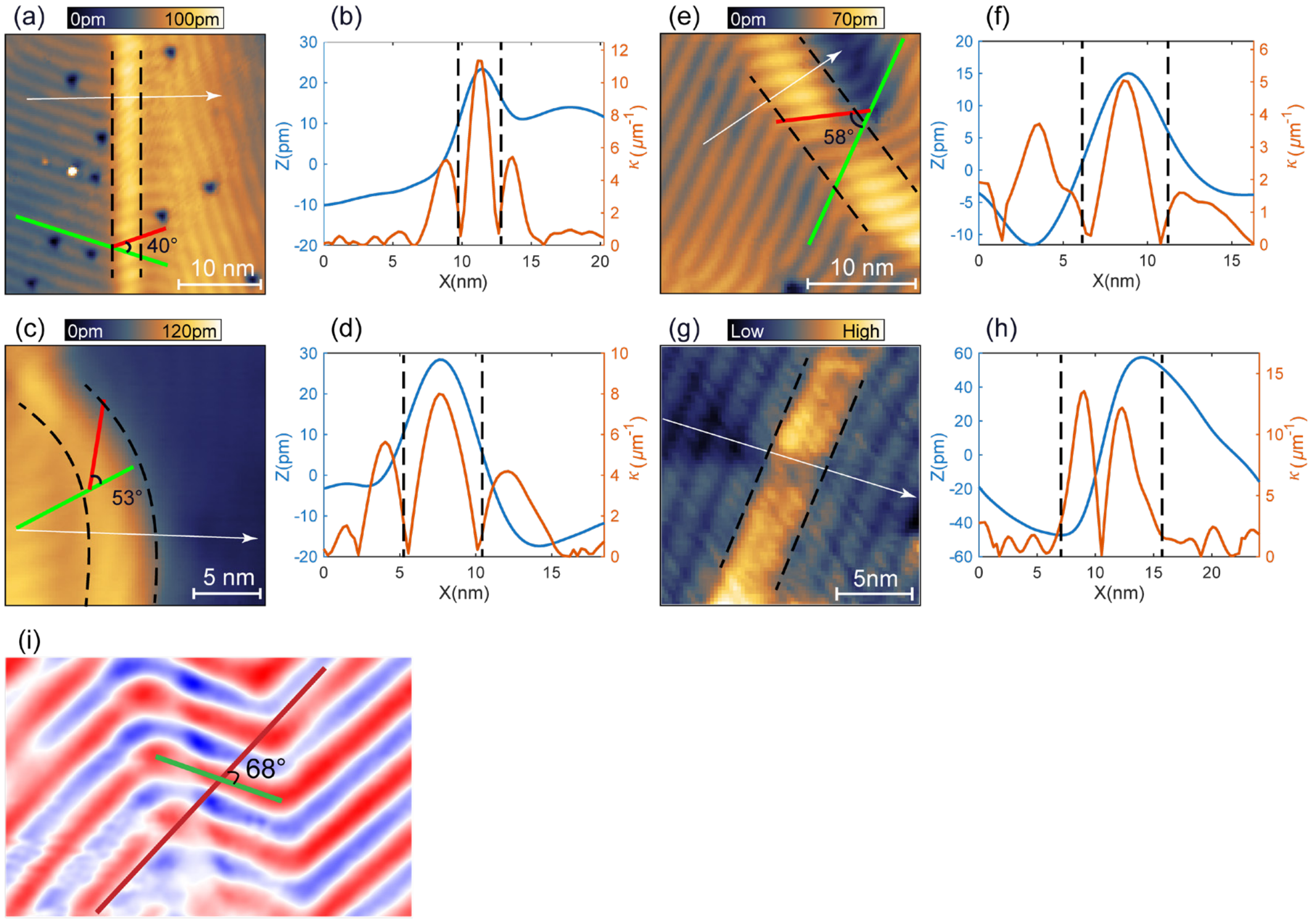

**Fig. S5. Deflections of spin spirals at locally curved surface.** (**a,e,c,g**) Zoom-in images of locally curved surface, the deflected angles of spin spirals are indicated. (**b,f,d,h**) The corresponding line profiles taken across the curved regions (along the arrows in the images). Blue curves are apparent height $Z$ (pm); orange curves are local curvature $\kappa$ (μm$^{-1}$). The curvature is peaked at the center of the curved region. (**i**) The $S_Z$ map of a deflected spin spiral.

## I-5: Additional STM data about few-layer $NiI_2$ films grown on $NbSe_2$

We have performed STM measurement on the $NiI_2$ film grown on $NbSe_2$ and compared the results with the $NiI_2$ film on HOPG. As shown in Fig. S6 below, the d$I$/d$V$ spectra of 1–2ML $NiI_2$ on $NbSe_2$ still exhibit an insulating gap, but its conduction band edge is shifted to positive bias by ~1.0 eV relative to 1ML $NiI_2$/HOPG. This substantial shift indicates a pronounced hole-doping effect in $NiI_2$ film on $NbSe_2$. Importantly, despite this large difference in charge transfer, we still observed a clear 2$\boldsymbol{q}$ charge modulation in 1ML and 2ML $NiI_2$ films on $NbSe_2$, as shown in Fig. S6b and c). The lateral period of this charge modulation is very close to that observed in 1ML/2ML $NiI_2$ on HOPG (Fig. 2d of the main text), therefore it indicates the spin-spiral order remains robust in $NiI_2$/$NbSe_2$. We further find that the orientation of q in $NiI_2$/$NbSe_2$ also forms a small angle with respect to [110] direction, which is ~ 9° for 1ML and ~12° for 2ML. These values are close to that of 1ML/2ML $NiI_2$ film on HOPG (Fig. 3h).

The data of $NiI_2$/$NbSe_2$ suggest that although the substrate can induce substantial charge transfer, the spin-spiral order and its wavevector are only weakly affected.

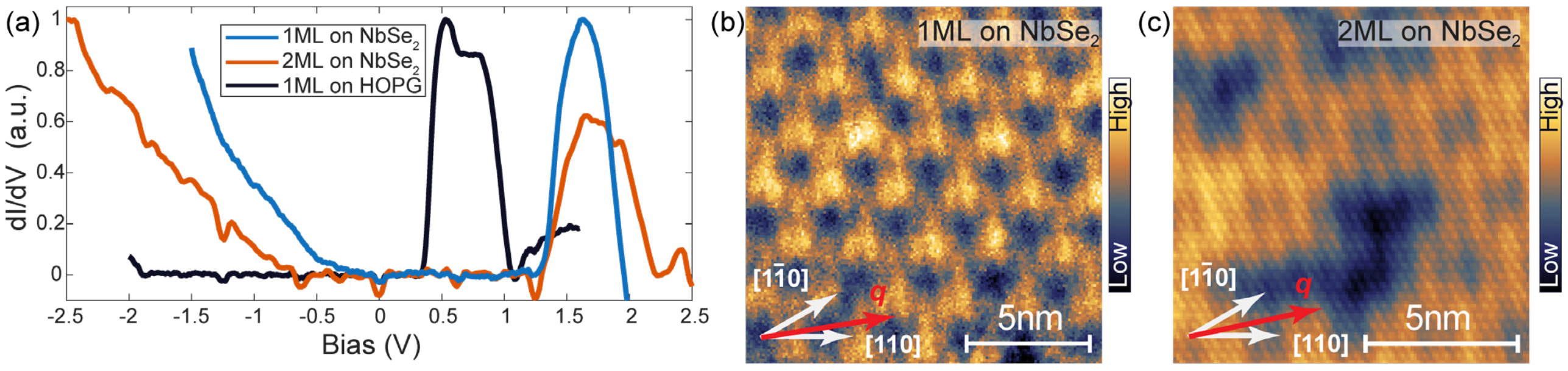

**Fig. S6. d*I*/d*V* spectra and 2*q* modulation of the $NiI_2$ film on $NbSe_2$ substrate.** (**a**) d*I*/d*V* spectra of $NiI_2$ film grown on different substrates, showing different interface charge transfer. (**b, c**) 2***q*** modulation on 1ML and 2ML $NiI_2$ film grown on $NbSe_2$ substrate, therefore the corresponding 1***q*** wavevector can be estimated (1ML: $\boldsymbol{\lambda_q}$=1.74nm, $\boldsymbol{\alpha}$=9°; 2ML: $\boldsymbol{\lambda_q}$=1.89nm, $\boldsymbol{\alpha}$=12°).

# Part II: Additional details about analytical calculation

## II-1: Magnetic parameters for solving the spin Hamiltonian.

The intralayer interaction parameters ($J_1 - J_3$, $K$, and $B$) with $U$ = 4 eV were adopted from the previous study [4]. The $J_3$ was enhanced from 2.25 meV to 2.63 meV to account for the deviation of spin-spiral propagation in monolayer $NiI_2$, following the solutions in the previous study [3]. The interlayer exchange parameters are extracted by DFT calculation (Methods). Due to discrepancies in DFT calculation parameters between the intralayer and interlayer studies, we fixed the ratios of the interlayer interactions and scaled their magnitudes to align with experimental data.

We also performed additional calculations to examine the dependence of exchange parameters on different $U_{\mathrm{eff}}$ values (2, 4, 6 eV). The results are summarized in Table S1 below. As expected for DFT+U calculation, the absolute values of exchange parameters show some dependence $U_{\mathrm{eff}}$, However, some key dimensionless ratios, such as $J_2/J_1$, $J_3/J_1$, which are more relevant to the noncollinear magnetic ground state of $NiI_2$, vary only weakly over this range. Meanwhile, the changes of the intralayer parameters $J_2^{\perp}$ and $J_{3,2}^{\perp}$ are also small. This suggest that the qualitative magnetic behavior of $NiI_2$ film is not sensitive to specific $U_{\mathrm{eff}}$. The thickness dependence of spin-spiral wavevector and a dimensional crossover behavior likely remains over a considerable range of $U_{\mathrm{eff}}$. We therefore chose $U_{\mathrm{eff}}$ = 4 eV since it provides a good agreement with experimental data.

**Table S1.** Dependence of the exchange parameters on the $U_{eff}$ value. The unit of $U_{eff}$ is eV, and the exchange parameters are given in meV. Note that, for the results with $U_{eff}$ = 4 eV, only the interlayer interactions were calculated and then adjusted using a fixed scaling factor, whereas the intralayer interactions were taken from the HSE-functional calculations reported in a previous work of $NiI_2$ [4].

| $U_{eff}$ | $J_1$ | $J_2/J_1$ | $J_3/J_1$ | $K$ | $B$ | $J_2^{\perp}$ | $J_{3,2}^{\perp}$ |
|---|---|---|---|---|---|---|---|
| 2 | -5.552 | 0.028 | -0.451 | 2.410 | -0.928 | 0.515 | 1.992 |
| *4 | -4.976 | 0.031 | -0.456 | 0.858 | -0.719 | 0.150 | 0.600 |
| 6 | -2.812 | 0.030 | -0.353 | 0.820 | -0.539 | 0.209 | 0.743 |

## II-2: Analytical energy derivation for the single-q state

To compute the energy of a certain spin configuration, we start with expanding the spin configuration $\{\boldsymbol{S_R}\}$ into normal modes, $\boldsymbol{S_R} = \sum_{\boldsymbol{k}} \boldsymbol{A_k} e^{i\boldsymbol{k}\cdot\boldsymbol{R}}$, where $\boldsymbol{k}$ is the wavevector in the first Brillouin zone.

The normal modes satisfy $\boldsymbol{A_k} = \boldsymbol{A}_{-\boldsymbol{k}}^*$ and orthogonality relations: $\sum_{\boldsymbol{R}} e^{i\boldsymbol{k}\cdot\boldsymbol{R}} e^{-i\boldsymbol{k}'\cdot\boldsymbol{R}} = N\delta_{\boldsymbol{k},\boldsymbol{k}'}$. The spins are normalized, i.e. $|\boldsymbol{S_R}|^2 = 1$, which means that the normal modes should satisfy the constrain: $\sum_{\boldsymbol{k},\boldsymbol{k}'} \boldsymbol{A}_{\boldsymbol{k}}^{\dagger} \boldsymbol{A}_{\boldsymbol{k}'} e^{-i(\boldsymbol{k}-\boldsymbol{k}')\cdot\boldsymbol{R}} = 1$. Considering second-order anisotropic magnetic interactions: $E = \frac{1}{2}\sum_{\boldsymbol{R},\boldsymbol{R}'} \boldsymbol{S}_{\boldsymbol{R}}^T \mathbb{J}_{\boldsymbol{R},\boldsymbol{R}'} \boldsymbol{S}_{\boldsymbol{R}'}$ where $\mathbb{J}_{\boldsymbol{R},\boldsymbol{R}'}$ is the interaction matrix with lattice translational symmetry $\mathbb{J}_{\boldsymbol{R},\boldsymbol{R}'} = \mathbb{J}_{\boldsymbol{R}-\boldsymbol{R}'',\boldsymbol{R}'-\boldsymbol{R}''}$. We can express the energy in terms of the normal mode amplitudes

$$E = \frac{N}{2}\sum_{\boldsymbol{k}} \boldsymbol{A}_{\boldsymbol{k}}^{\dagger} \mathbb{M}_{\boldsymbol{k}} \boldsymbol{A}_{\boldsymbol{k}}\,, where\ \mathbb{M}_{\boldsymbol{k}} = \sum_{\boldsymbol{R}} \mathbb{J}_{0,\boldsymbol{R}} e^{i\boldsymbol{k}\cdot\boldsymbol{R}}$$

Then we consider a single-$\boldsymbol{q}$ spiral state, which has only 2 normal modes, $\pm\boldsymbol{k}$. The energy and constrain can be simplified as

$$\frac{E}{N} = \boldsymbol{A}_{\boldsymbol{k}}^{\dagger} \mathbb{M}_{\boldsymbol{k}} \boldsymbol{A}_{\boldsymbol{k}}, \qquad \boldsymbol{A}_{\boldsymbol{k}} \cdot \boldsymbol{A}_{\boldsymbol{k}}^* = \frac{1}{2}\ and\ \boldsymbol{A}_{\boldsymbol{k}}^2 = 0.$$

Note that there is no DMI in $NiI_2$, so the matrix $\mathbb{M}_{\boldsymbol{k}}$ is symmetric and an orthogonal basis can be found to diagonalize $\mathbb{M}_{\boldsymbol{k}}$. Assuming the eigenvalues of $\mathbb{M}_{\boldsymbol{k}}$ are $\lambda_{\boldsymbol{k},1} < \lambda_{\boldsymbol{k},2} < \lambda_{\boldsymbol{k},3}$, while the corresponding coordinates of $\boldsymbol{A}_{\boldsymbol{k}}$ is $(a_1, a_2, a_3)^{\mathrm{T}} + i(b_1, b_2, b_3)^{\mathrm{T}}$, then the problem to be solved becomes

$$\frac{E}{N} = \lambda_{\boldsymbol{k},1}(a_1^2 + b_1^2) + \lambda_{\boldsymbol{k},2}(a_2^2 + b_2^2) + \lambda_{\boldsymbol{k},3}(a_3^2 + b_3^2)$$

$$a_1^2 + a_2^2 + a_3^2 = \frac{1}{4}, b_1^2 + b_2^2 + b_3^2 = \frac{1}{4}, a_1 b_1 + a_2 b_2 + a_3 b_3 = 0.$$

To find the ground spiral state, the energy density $E/N$ should be minimized under the constrains. A simple insight suggests that $a_3^2 + b_3^2 = 0$, i.e., $a_3 = b_3 = 0$, which exactly gives the minimum value

$$\frac{E}{N} = \frac{\lambda_{\boldsymbol{k},1} + \lambda_{\boldsymbol{k},2}}{4}.$$

Then minimizing the energy by $\boldsymbol{k}$ gives the propagation direction and the period. For the Hamiltonian we studied, the monolayer minimal energy density is expressed as

$$\frac{E}{N} = \left(J_1 + \frac{K}{2}\right)\left[cos(k_x a) + cos\left(\frac{k_x a}{2} - \frac{\sqrt{3}k_y a}{2}\right) + cos\left(\frac{k_x a}{2} + \frac{\sqrt{3}k_y a}{2}\right)\right]$$
$$+J_2\left[cos(\sqrt{3}k_y a) + cos\left(\frac{3k_x a}{2} - \frac{\sqrt{3}k_y a}{2}\right) + cos\left(\frac{3k_x a}{2} + \frac{\sqrt{3}k_y a}{2}\right)\right]$$
$$+\left(J_3 + \frac{B}{2}\right)\left[cos(2k_x a) + cos\left(k_x a - \sqrt{3}k_y a\right) + cos\left(k_x a + \sqrt{3}k_y a\right)\right] + \frac{3}{2}B$$
$$-\frac{K}{2}max\left\{cos(k_x a)\,, cos\left(\frac{k_x a}{2} - \frac{\sqrt{3}k_y a}{2}\right), cos\left(\frac{k_x a}{2} + \frac{\sqrt{3}k_y a}{2}\right)\right\},$$

where $k_x$ is the component of the vector $\boldsymbol{k}$ along the [110] direction, and $k_y$ is in the direction $[1\bar{1}0]$.

**II-3: Energy contribution of biquadratic interaction for single-q state**

For a single-$\boldsymbol{q}$ spiral, the spins can be expanded into normal modes, $\boldsymbol{S_R} = \boldsymbol{A_k}e^{i\boldsymbol{k}\cdot\boldsymbol{R}} + \boldsymbol{A_k^*}e^{-i\boldsymbol{k}\cdot\boldsymbol{R}}$. Substituting it into the expression for the biquadratic interaction energy yields

$$E_B = \frac{1}{2}B\sum_{\boldsymbol{R},\boldsymbol{R}'}(\boldsymbol{S_R}\cdot\boldsymbol{S_{R'}})^2$$
$$= \frac{1}{2}B\sum_{\boldsymbol{R},\boldsymbol{R}'}\left[\left(\boldsymbol{A_k}e^{i\boldsymbol{k}\cdot\boldsymbol{R}} + \boldsymbol{A_k^*}e^{-i\boldsymbol{k}\cdot\boldsymbol{R}}\right)\cdot\left(\boldsymbol{A_k}e^{i\boldsymbol{k}\cdot\boldsymbol{R}'} + \boldsymbol{A_k^*}e^{-i\boldsymbol{k}\cdot\boldsymbol{R}'}\right)\right]^2$$
$$= \frac{1}{2}B\sum_{\boldsymbol{R},\boldsymbol{R}'}\left\{\boldsymbol{A_k^2}e^{i\boldsymbol{k}\cdot(\boldsymbol{R}+\boldsymbol{R}')} + \boldsymbol{A_k^*}^2e^{-i\boldsymbol{k}\cdot(\boldsymbol{R}+\boldsymbol{R}')} + 2\boldsymbol{A_k}\cdot\boldsymbol{A_k^*}\,cos[\boldsymbol{k}\cdot(\boldsymbol{R}-\boldsymbol{R}')]\right\}^2,$$

where $\boldsymbol{R}'$ stands for the first nearest-neighbors of $\boldsymbol{R}$.

The spins are normalized, i.e. $|\boldsymbol{S_R}|^2 = 1$, corresponding to the normal modes satisfying the constrain: $\boldsymbol{A_k}\cdot\boldsymbol{A_k^*} + \mathrm{Re}\left(\boldsymbol{A_k^2}e^{2i\boldsymbol{k}\cdot\boldsymbol{R}}\right) = \frac{1}{2}$. This constrain should be satisfied by all $\boldsymbol{R}$, thus we have $\boldsymbol{A_k}\cdot\boldsymbol{A_k^*} = \frac{1}{2}$ and $\boldsymbol{A_k^2} = 0$. Then $E_B$ can be simplified as

$$E_B = \frac{1}{2}B\sum_{\boldsymbol{R},\boldsymbol{R}'}cos^2[\boldsymbol{k}\cdot(\boldsymbol{R}-\boldsymbol{R}')] = \frac{N}{4}B\sum_{\boldsymbol{R}_1}[cos(2\boldsymbol{k}\cdot\boldsymbol{R}_1) + 1].$$

where $\boldsymbol{R}_1$ indicates the vectors of the first nearest-neighbors. Note that the energy contribution of the third-neighbor coupling $J_3$ is $E_3 = \frac{N}{2}J_3\sum_{\boldsymbol{R}_3}\cos(\boldsymbol{k}\cdot\boldsymbol{R}_3)$, and for the triangular lattice we considered, $\boldsymbol{R}_3 = 2\boldsymbol{R}_1$. Therefore, the energy contribution of biquadratic interaction is equivalent to a $J_{3,\text{eff}} = B/2$, differing only by a constant term.

**II-4: Definition of different phase of spin spiral in phase diagram**

The magnetic ground states shown in Fig. 4b and c of the main text can be classified by the location of the corresponding single ordering vector $\boldsymbol{q}$ within an irreducible wedge of the first Brillouin zone, as illustrated in Fig. S7a. In this representation, each magnetic phase is associated with a distinct region in reciprocal space, and the evolution of the real-space spin texture can be tracked by $\boldsymbol{q}$.

The $\boldsymbol{q}$ = 0 point (Γ point) corresponds to a FM state. When $\boldsymbol{q}$ reaches the M point, the spins are aligned within rows along $[110]$ and reverse between neighboring rows along $[1\bar{1}0]$, corresponding to a commensurate stripe AFM state. All other phases in our calculation are incommensurate, and can be further divided depending on $\boldsymbol{q}$. When $\boldsymbol{q}$ is on $\overrightarrow{\Gamma K}$ (blue segment along $[110]$ direction), the magnetic state is denoted as $\text{IC}^{[110]}$. In our calculation, no stable ground state belonging to this phase is found on the red segment in our calculations. When $\boldsymbol{q}$ is on $\overrightarrow{\Gamma M}$ (yellow line along $[1\bar{1}0]$ direction), the magnetic state is denoted as $\text{IC}^{[1\bar{1}0]}$. When $\boldsymbol{q}$ is located inside the color-coded triangle region, the magnetic state is denoted as $\text{IC}^{\alpha}$ phase since there is a finite angle $\alpha$ between $\boldsymbol{q}$ and $[110]$ direction, and the color scale in the triangle represents the angle $\alpha$. A further incommensurate state appears when $\boldsymbol{q}$ is on $\overrightarrow{MK}$ (green line). This segment is symmetry-equivalent to $\overrightarrow{KM^{(2)}}$ (green dash line, along $[110]$ direction). Therefore, the corresponding magnetic state is denoted as a second $\text{IC}^{[110]}$ phase, with different wavelength relative to the former one.

**II-5: Effective intralayer couplings of the interlayer interactions**

As an illustration of the effective mapping between intralayer and interlayer couplings, we consider the interlayer second-neighbor coupling $J_2^{\perp}$ as an example, with its corresponding neighboring vectors depicted in Fig. S7b. The site at $\boldsymbol{R}_0$ contributes an energy term given by $J_2^{\perp} \sum_{\boldsymbol{R}_{2\perp}} \boldsymbol{S}_{\boldsymbol{R}_0} \cdot \boldsymbol{S}_{\boldsymbol{R}_0+\boldsymbol{R}_{2\perp}}$, where $\boldsymbol{R}_{2\perp}$ denotes the neighboring vectors associated with $J_2^{\perp}$. Experimentally, the spin alignment between adjacent layers is found to be AFM, implying $\boldsymbol{S}_{\boldsymbol{R}_0} = -\boldsymbol{S}_{\boldsymbol{R}_{0\perp}}$ Consequently, the energy expression reduces to $-J_2^{\perp} \sum_{\boldsymbol{R}_{2\perp}} \boldsymbol{S}_{\boldsymbol{R}_{0\perp}} \cdot \boldsymbol{S}_{\boldsymbol{R}_0+\boldsymbol{R}_{2\perp}}$. Interestingly, this energy expression corresponds to the contribution from the intralayer first-neighbor interaction, given by $\frac{J_{1,\text{eff}}}{2} \sum_{\boldsymbol{R}_1} \boldsymbol{S}_{\boldsymbol{R}_{0\perp}} \cdot \boldsymbol{S}_{\boldsymbol{R}_{0\perp}+\boldsymbol{R}_1}$, where $\boldsymbol{R}_1$ represents the neighboring vectors corresponding to $J_1$, and $J_{1,\text{eff}} = -2J_2^{\perp}$. For an $N$-layer system, there are only $N-1$ interlayer couplings; therefore, the average effective intralayer coupling per layer is given by $J_{1,\text{eff}} = -\frac{2(N-1)}{N} J_2^{\perp}$. Consequently, the interlayer interactions in the multilayer system are effectively represented by an equivalent intralayer term. Following the same approach, the interlayer third-neighbor interaction can be analyzed, where iodine-mediated symmetry breaking distinguishes two coupling types that contribute equally to the corresponding intralayer second-neighbor term.

If the spin alignment between adjacent layers is not strictly antiparallel, we can introduce a uniform phase difference $\phi$ between successive layers, consistent with the behavior in the bulk system. In this case, the spin components in adjacent layers are related by $\boldsymbol{S}_{\boldsymbol{R}_{0\perp}} = \boldsymbol{S}_{\boldsymbol{R}_0} \cos\phi$, leading to an effective intralayer coupling given by $J_{1,\mathrm{eff}} = \frac{2(N-1)}{N} J_2^{\perp} \cos\phi$. The conclusion that AFM interlayer interactions can be conceptually mapped onto effective FM intralayer couplings remains valid when $\cos\phi < 0$.

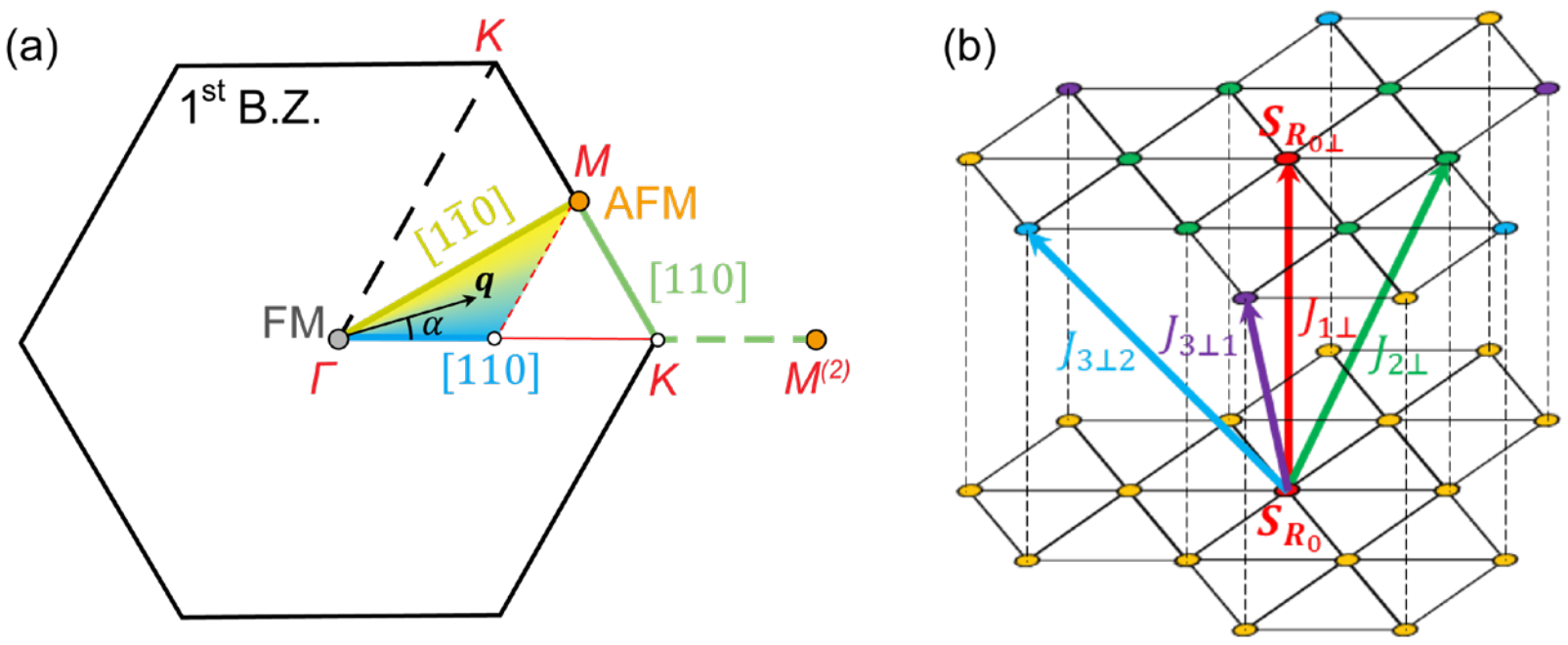

**Fig. S7. Definition of magnetic states and interlayer couplings.** (**a**) Definition of different magnetic states in reciprocal space. (**b**) Schematic image of the neighbors of interlayer couplings for AA stacking.

### II-6: Magnetic parameters under different stacking configuration

To evaluate whether the stacking registry qualitatively changes the magnetic interactions relevant to the spin-spiral state, we performed DFT calculations for a bilayer $NiI_2$ which contains AA, AB, and AC stacking regions, as summarized in Fig. S8 and Table S2 below. The calculations show that AA and AB stackings have similar intralayer exchange parameters, AFM-favored spin alignment (Fig. S8c) and interlayer distances (Fig. S8d). Besides, the dominant interlayer exchange parameters of them are both AFM (Table S2). Therefore, AA and AB stacking are expected to produce quantitatively different but qualitatively similar magnetic behavior for the spin-spiral physics discussed here. In contrast, AC stacking shows a different dominant interlayer exchange parameter ($J_{11}$), indicating that stacking can still tune the magnetic interactions. Nevertheless, AC stacking still favors antiferromagnetic interlayer alignment (Fig. S8c).

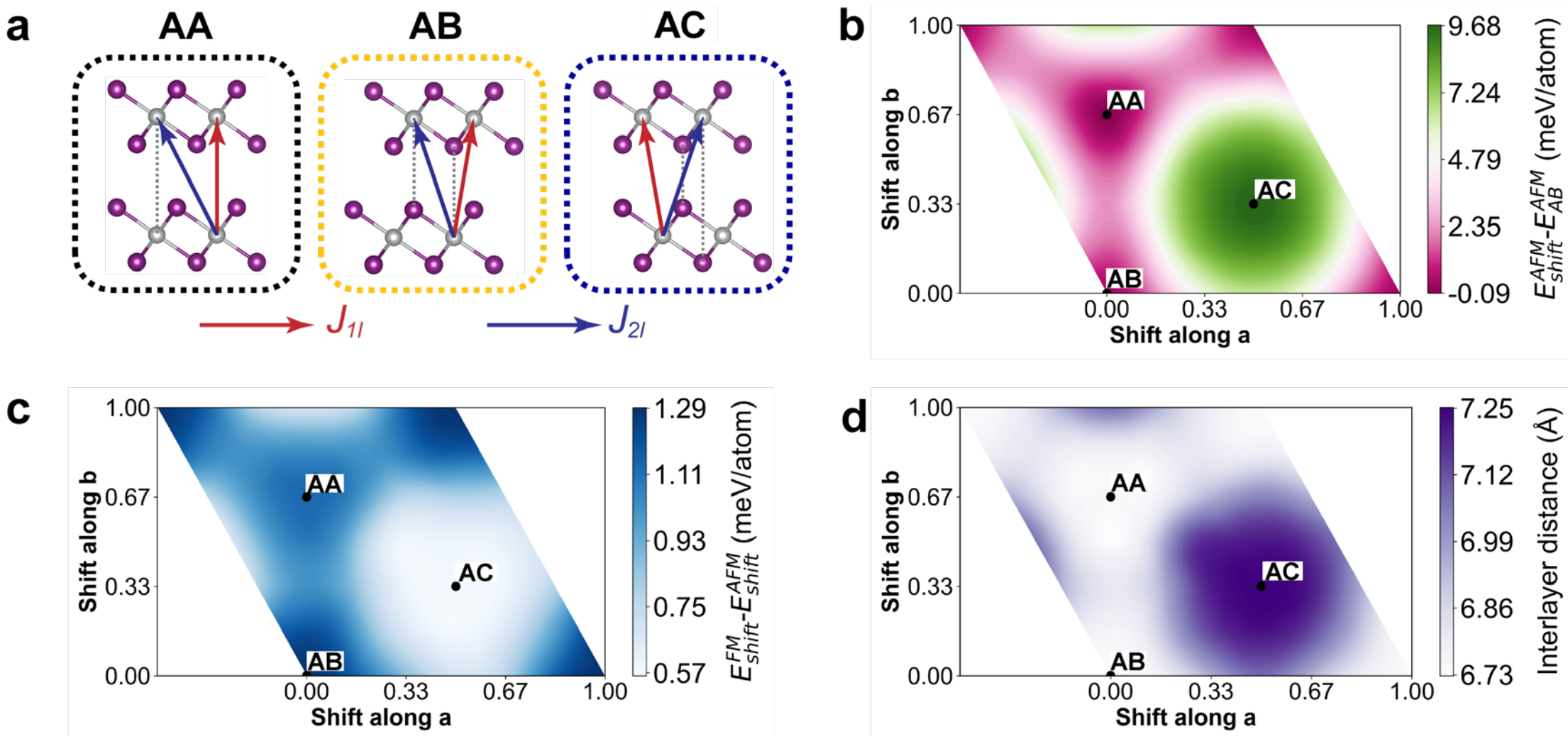

**Fig. S8. Stacking-dependent structural and magnetic properties of twist-free bilayer $NiI_2$.** (**a**) Schematic structures of AA, AB, and AC stackings. Representative interlayer exchange paths up to 2nd-nearest neighbor ($J_{1l}$ and $J_{2l}$) are indicated. (**b**) Relative stacking energy $\Delta E = E^{AFM}_{shift} - E^{AFM}_{AB}$ as a function of in-plane layer shift. (**c**) Interlayer magnetic energy difference $E^{FM}_{shift} - E^{AFM}_{shift}$. Positive values indicate that AFM interlayer alignment is favored. (**d**) Interlayer distance as a function of stacking registry.

**Table S2.** DFT four-state magnetic interaction parameters for twist-free bilayer $NiI_2$ with AA, AB, and AC stacking. The parameters include the single-ion anisotropy (SIA), 1st and 3rd intralayer exchanges ($J_1$, $J_3$), the ratio $|J_3/J_1|$, and 1st–2nd interlayer couplings ($J_{1l}$ and $J_{2l}$). Units are meV.

| Parameter | AA | AB | AC |
|---|---|---|---|
| SIA | 0.152 | 0.137 | 0.157 |
| $J_1$ | -3.547 | -3.547 | -3.593 |
| $J_3$ | 2.746 | 2.856 | 2.964 |
| $\lvert J_3/ J_1\rvert$ | 0.774 | 0.805 | 0.825 |
| $J_{1l}$ | 0.004 | -0.057 | 0.477 |
| $J_{2l}$ | 0.152 | 0.888 | -0.013 |

## II-7: Néel temperature in the multilayer $NiI_2$

We have performed Monte Carlo simulations of the temperature-dependent specific heat for 1–5 ML $NiI_2$, as well as the AA-stacked bulk limit. The specific heat is proportional to the energy fluctuation, which is strongly enhanced at phase transition point. Therefore, the peak position of the specific heat is used to estimate the phase transition temperature, namely the Néel temperature. Here in our system, the phase transition corresponds to a magnetic ordering transition from an intralayer spiral and interlayer antiferromagnetic spin structure to a paramagnetic state. The results are shown in Fig. S9, which show that as the thickness increases from 1–5 ML, the specific heat peak, i.e., the Néel temperature, gradually shifts to higher temperature. This indicates that the antiferromagnetic interactions within the system are progressively enhanced as the number of layers increases, suggesting an antiferromagnetic character of the interlayer coupling, which is consistent with our analytical model. This trend of Néel temperature is also consistent with the experimentally observed thickness-dependent magnetic behavior [5, 6].

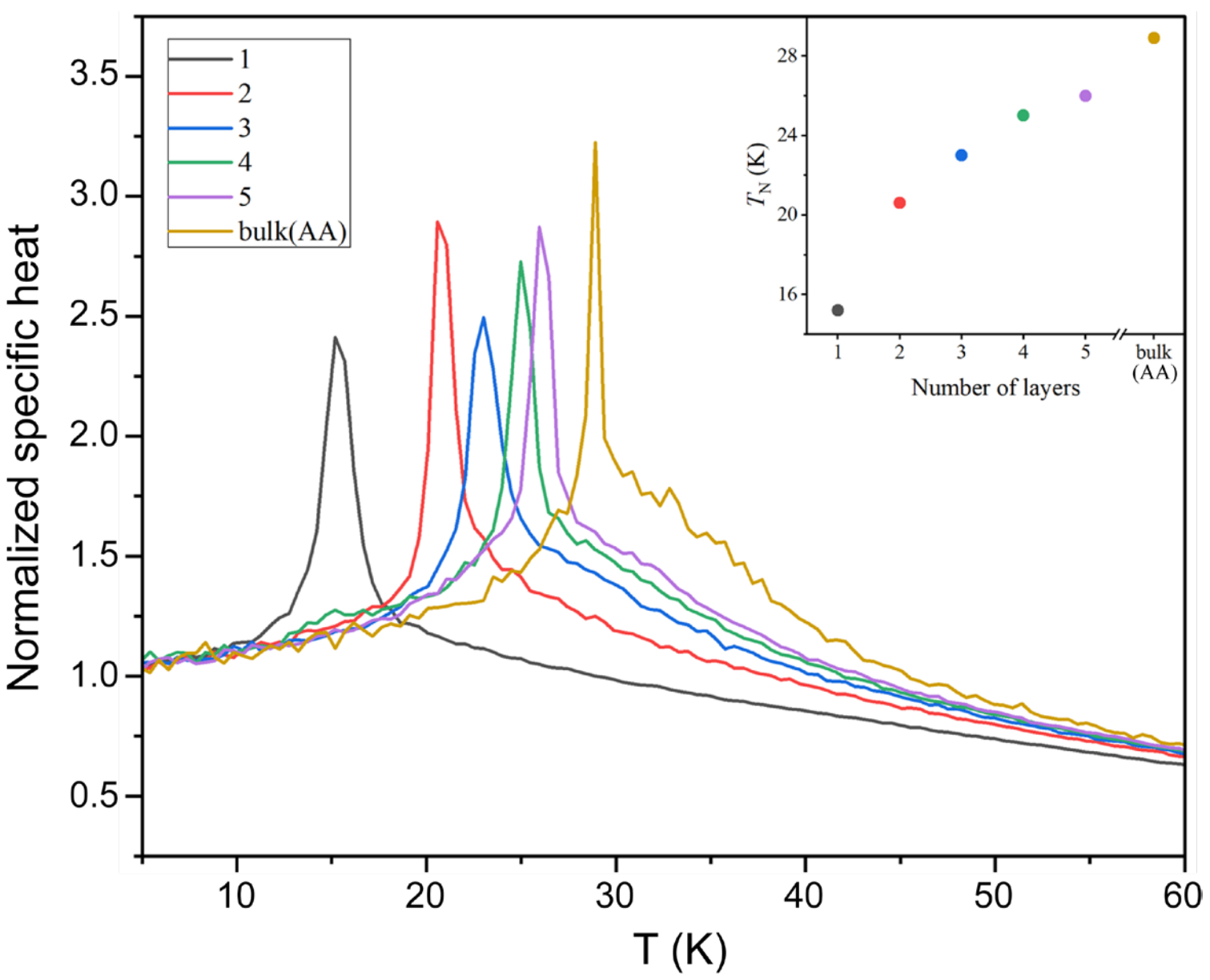

**Fig. S9. Temperature dependence of the specific heat obtained from Monte Carlo simulations.** The specific heat is normalized by the specific heat at the lowest temperature. Inset: Néel temperature $T_N$ of 1–5 ML and the AA-stacked bulk limit $NiI_2$ extracted from peak positions of specific heat.

## Part III: Additional details about Spin–lattice dynamics simulations.

### III-1: Dataset preparation for SpinGNN++

We build a spin-lattice dataset using DFT for the learning of neural network. The static DFT dataset was assembled for monolayer $NiI_2$ by jointly sampling atomic geometries and magnetic configurations in a balanced manner. Structural snapshots were taken from a finite–temperature ab initio molecular-dynamics (MD) trajectory and subsampled at a fixed stride; 50 perturbed configurations were retained. Unless stated otherwise, single-point calculations used the same electronic settings as described in Methods in the main text. The data were split into training, validation and test with an 8:1:1 ratio. For model training, a radial cutoff of 9.0 Å was adopted to include in-plane interactions up to approximately the third neighbor.

**Sampling procedure.** For each selected MD structure, spin states were generated on the uniform angle grid $\Theta = \{0°, 36°, \ldots, 324°\}$ (10 values). Define

$$\widehat{\boldsymbol{e}_{xy}}(\theta) = (cos\,\theta\,, sin\,\theta\,, 0), \qquad \widehat{\boldsymbol{e}_{yz}}(\theta) = (0, cos\,\theta\,, sin\,\theta),$$

and let $\mathbf{m}_i^{(0)}$ denote the target moment on site $i$ with fixed magnitude. The following families were included:

- **IFM**$(\theta)$: in-plane ferromagnet, $\mathbf{m}_i^{(0)} \parallel \widehat{\boldsymbol{e}_{xy}}(\theta)$; 10 angles.
- **OFM**$(\theta)$: out-of-plane ferromagnet, $\mathbf{m}_i^{(0)} \parallel \widehat{\boldsymbol{e}_{yz}}(\theta)$; 10 angles.
- **IAABB**$(\theta)$: in-plane AABB antiferromagnet with common axis $\widehat{e_{xy}}(\theta)$; 10 angles.
- **OAABB**$(\theta)$: out-of-plane AABB antiferromagnet with common axis $\widehat{e_{yz}}(\theta)$; 10 angles.
- **rand**: noncollinear states with independent random unit directions per magnetic site (fixed magnitude) ; 40 states.

Thus, each structural frame contributes 80 spin configurations (10+10+10+10+40). With N = 50 frames this gives 80 × 50 = 4000 targeted single-point calculations; after enforcing strict electronic-convergence criteria and removing pathological outliers, **3696** calculations remained and were used in the final dataset.

### III-2: Performance of Spin-Allegro++

We perform SpinGNN++ framework in its Allegro implementation, Spin-Allegro++[7, 8]. Spin-Allegro++ is employed with two layers of a multitask spin-equivariant neural network (MSENN) and a time-reversal-equivariant neural network (TENN) to fit the dataset, achieving a mean absolute error (MAE) of 0.036 meV/atom and an $R^2$ score of 0.9999 on the test set. These results indicate that Spin-Allegro++ accurately captures higher-order spin-lattice interactions.

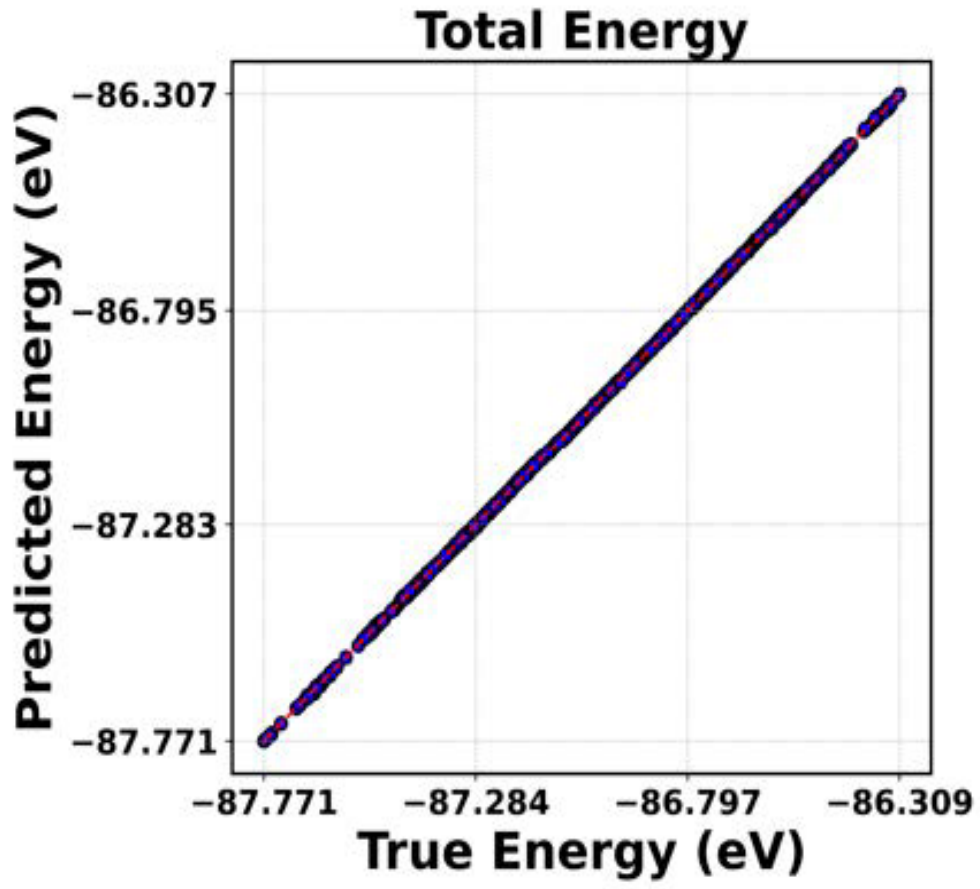

**Fig. S10. Performance of Spin-Allegro++ on monolayer $NiI_2$.**

The energy disparities between different magnetic states are calculated using both DFT and Spin-Allegro++, as shown in Table S3. The subtle energy differences and energy rankings are accurately reproduced by Spin-Allegro++, facilitating the precise prediction of the magnetic ground state and confirming the accuracy of Spin-Allegro++.

**Table S3:** The energy difference per atom between various magnetic states of monolayer $NiI_2$, calculated using first-principles calculations and Spin-Allegro++, is presented in meV/atom.

| | OFM | OAABB | OABAB | IFM | IAABB | IABAB |
|---|---|---|---|---|---|---|
| DFT | 0.000 | -0.924 | 5.414 | -0.016 | -1.085 | 4.958 |
| Net | 0.000 | -0.957 | 6.595 | 0.019 | -1.130 | 6.106 |

### III-3: The simulation of spin structure for curved $NiI_2$ lattice.

To simulate the spin structure of original and curved $NiI_2$ lattice, we first validate our SpinGNN++ potential by comparing its predictions of key magnetic parameters with results from DFT. The extracted magnetic parameters are as follows: the isotropic Heisenberg exchange $J_1$=−4.22 meV, the Kitaev interaction $K$=1.28 meV, the biquadratic term $B$=−0.79 meV, the AFM third-neighbor exchange $J_3$=2.64 meV, and the single–ion anisotropy $A_z$=0.137 meV which favors in-plane spin alignment. These parameters are consistent with the PBE-based DFT results reported before [4].

We then employed a $30 \times 30 \times 1$ supercell, constructed from the minimal rectangular unit cell, to optimize both atomic and spin degrees of freedom via annealing and conjugate-gradient methods. The machine learning potential reproduces the magnetic ground state of monolayer $NiI_2$, a [110] proper screw with $\boldsymbol{q}$= (0.22, 0.22, 0), consistent with experimental observations[5, 9-11].

To model curvature-induced effects, the initial atomic configuration was generated by symmetrically bending the relaxed monolayer structure of $NiI_2$ along the in-plane $x$ axis. The bending transformation was performed using a custom Python script based on the Atomic Simulation

Environment (ASE), in which the atomic coordinates were mapped onto a cylindrical surface with a bending radius of $R$ = 1000Å ($\kappa = 10\mu m^{-1}$). This operation preserves the in-plane lattice periodicity while introducing a gradual out-of-plane displacement corresponding to a symmetric curvature about the central plane of the layer. The resulting bent structure was subsequently used as the initial configuration for the spin–lattice dynamics simulations.

After spin–lattice dynamics relaxation, the bent lattice is only weakly modified, indicating that the curvature-induced deformation is mechanically stable. The maximum out-of-plane displacement along the $c$ direction reaches ~4.89 Å, showing that monolayer $NiI_2$ can accommodate substantial curvature without structural degradation. Analysis of the relaxed lattice geometry further shows that the local bond length and angles of top/bottom I-Ni bonds change most significantly in the curved region (Fig. S11a–d). This confirms a direct structural origin of spatial modulations of $J_3/J_1$ and $K$ discussed in the main text.

Additionally, we also calculated the topological charge density from the spin configuration in Fig. S11e. Here the definition of the topological charge is $Q = \int \rho_{topo}(x, y)dxdy$, where the topological charge density is $\rho_{topo}(x, y) = \frac{1}{4\pi}\boldsymbol{m} \cdot \left(\frac{\partial \boldsymbol{m}}{\partial x} \times \frac{\partial \boldsymbol{m}}{\partial y}\right)$. $m$ is the normalized magnetization. Here we present the topological charge distribution in Fig. S11f. Notably, no direct correlation is found between the topological charge and the local deflection. Non-zero topological charge only distribute on the upper side of the simulation area, which could due to some finite-size effects of the spin structure simulation. For comparison, it is reported that the spin-spiral domain wall of $NiI_2$ is topological nontrivial[3, 4]. This further indicates that the deflected spin-spiral at the wrinkle is not a domain wall, but instead a distinct spin texture.

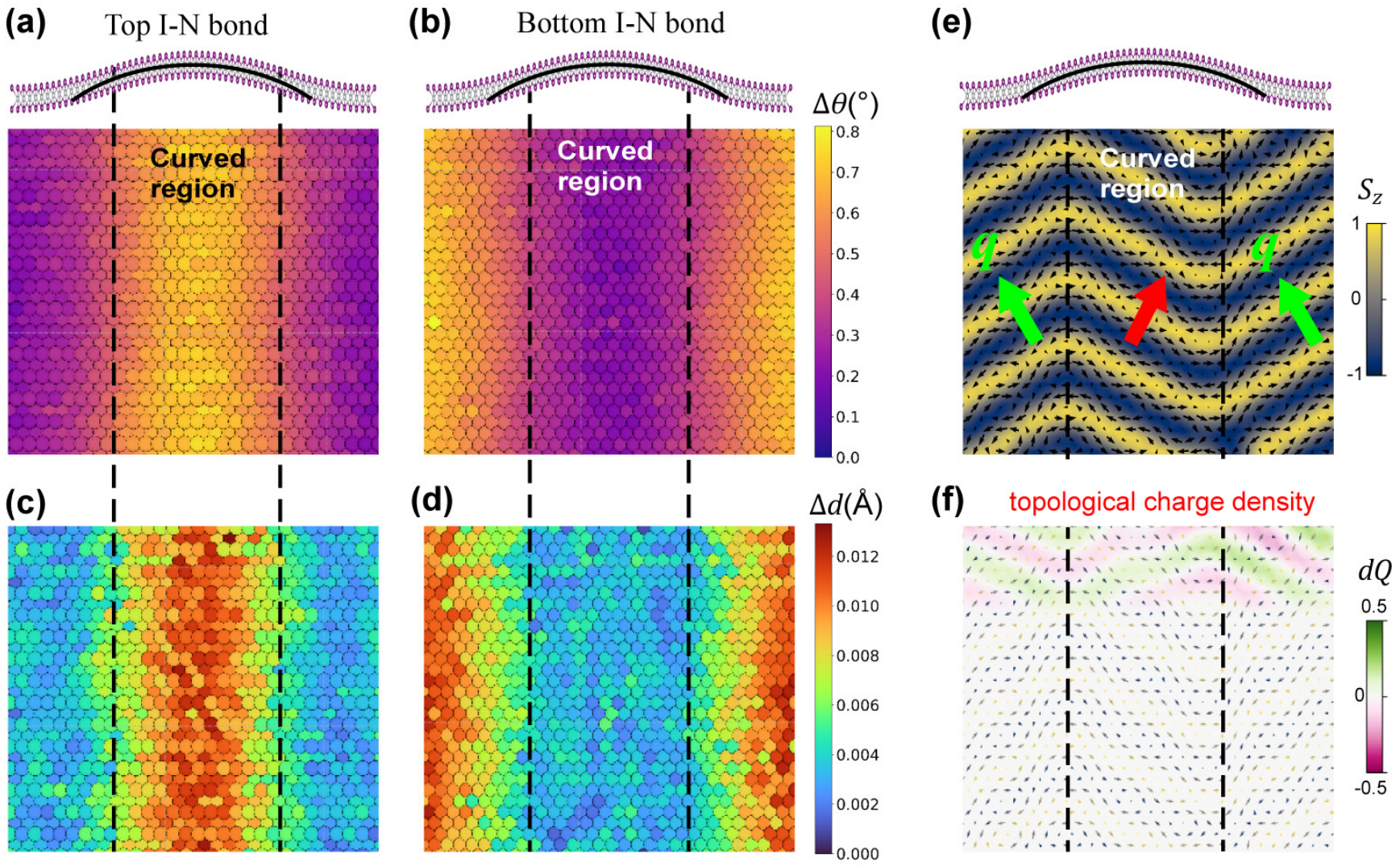

**Fig. S11. Additional data of lattice structure and spin structure obtained by SpinGNN++ simulation.** (**a, b**) Bond angle difference of top and bottom I-Ni bond (relative to that of flat lattice), respectively. (**c, d**) Bond length difference on top and bottom I-Ni bond, respectively. (**e, f**) simulated spin structure and the corresponding topological charge density.